\documentclass[useAMS,usenatbib]{mn2e}
\usepackage{graphicx}
\usepackage{natbib}
\usepackage{amsmath,amssymb,latexsym}
\usepackage{multirow}

\newcommand{\RE}{R_{\oplus}}
\newcommand{\ME}{M_{\oplus}}
\newcommand{\Tsurf}{$T_{\rm 1\,bar}$}

\newcommand{\Mjup}{\mbox{$M_{\rm J}$}}

\newcommand{\Tcore}{\mbox{$T_{\rm core}$}}
\newcommand{\Yatm}{\mbox{$Y_{\rm atm}$}}

\newcommand{\FLDD}{F_{\rm LDD}}

\newcommand{\Rinv}{R_0^{-1}}
\newcommand{\Rcrit}{R_{\rm crit}^{-1}}

\title[Jupiter: H/He phase separation and LDD convection]{An Exploration of Double Diffusive Convection in Jupiter as a Result of Hydrogen-Helium Phase Separation}

\author[Nettelmann et al.]{N.~Nettelmann$^1$
\thanks{E-mail: nadinen@ucolick.org (NN); jfortney@ucsc.edu (JJF)}, 
J.~J.~Fortney$^1$, K.~Moore$^{1,2}$, and C.~Mankovich$^1$\\
$^1$Department of Astronomy and Astrophysics, University of California, Santa Cruz, CA-95064, USA\\
$^2$Baskin School of Engineering, University of California, Santa Cruz, CA-95064, USA}

\begin{document}

\date{Accepted 2014 December 08. Received 2014 December 08; in original form 2014 July 17}

\pagerange{\pageref{firstpage}--\pageref{lastpage}} \pubyear{2014}

\maketitle

\label{firstpage}

\begin{abstract}
Jupiter's atmosphere has been observed to be depleted in helium ($\Yatm\sim 0.24$), suggesting active helium sedimentation in the interior. This is accounted for in standard Jupiter structure and evolution models through the assumption of an outer, He-depleted envelope that is separated from the He-enriched deep interior by a sharp boundary. Here we aim to develop a model for Jupiter's inhomogeneous thermal evolution that relies on a more self-consistent description of the internal profiles of He abundance, temperature, and heat flux. We make use of recent numerical simulations on H/He demixing, and on layered (LDD) and oscillatory (ODD) double diffusive convection, and assume an idealized planet model composed of a H/He envelope and a massive core. A general framework for the construction of interior models with He rain is described. Despite, or perhaps because of, our simplifications made we find that self-consistent models are rare. For instance, no model for ODD convection is found. We modify the H/He phase diagram of Lorenzen et al. to reproduce Jupiter's atmospheric helium abundance and examine evolution models as a function of the LDD layer height, from those that prolong Jupiter's cooling time to those that actually shorten it.  Resulting models that meet the luminosity  constraint have layer heights of $\approx 0.1$--1 km, corresponding to $\approx 10$,--20,000 layers in the rain zone between $\sim 1$ and 3--4.5 Mbars. Present limitations and directions for future work are discussed, such as the formation and sinking of He droplets.
\end{abstract}

\begin{keywords}
planets and satellites: individual(Jupiter), interiors, physical evolution -- convection
\end{keywords}

\section{Introduction}

Helium abundance measurements in Jupiter's atmosphere, beginning with ground based, aircraft, and \emph{Pioneer} 10,11 spacecraft observations and culminating in the \emph{Galileo Entry Probe} experiment, exhibit a remarkable agreement about a depletion in helium compared to the protosolar value \citep{OrtonIng76,Gautier81,Zahn98,Niemann98}. The \emph{Galileo} in-situ measurement also revealed a significantly sub-protosolar neon abundance. Both the He and Ne abundances are thought to result from phase separation of helium from hydrogen under high pressures and of downward rain of He-Ne rich droplets \citep{Stev98,WilMil10}. 

A helium rain region in Jovian planets, long predicted to occur \citep{Salpeter73,Stevenson75} is likely accompanied by a composition gradient and superadiabatic temperatures \citep{SS77b}. The precious in-situ observation of Jupiter's atmospheric He abundance of $Y:=M_{\rm He}/(M_{\rm H}+M_{\rm He})=0.238 \pm 0.005$ by mass \citep{Zahn98} thus indicates a more exotic Jovian interior than so far described by standard models. Those commonly represent Jupiter by few, sharply separated homogeneous and adiabatic layers \citep{Chabrier+92,Saumon+92,Guillot+97,GZ99,SG04,N+08,N+12}, even if He rain is explictly accounted for in the planet's thermal evolution \citep{Hubbard99}. 
These simplifications have of course been quite valid, as neither H/He phase diagrams with predictive power  existed, 
nor was a theory for heat transport in an inhomogeneous medium under Jovian interior conditions 
available.  Both are important, but not necessarily sufficient, for determining the gradients in 
He abundance and in temperature in Jupiter's interior. 

Thanks to growing computer power, this situation has changed in recent years.
Using ab initio simulations, \citet{Morales09,Lorenzen09,Lorenzen11,Morales13} have studied the demixing behaviour of He from H under Jupiter and Saturn interior conditions, where hydrogen undergoes a transition from non-metallic to metallic fluid. 
Semi-convection, a fluid instability that can occur in the presence of a destabilizing temperature gradient and a  stabilizing composition gradient, has recently been investigated by  \citet{Rosenblum11,Mirouh12,Wood13} using 3D-numerical simulations.  They observe semi-convection, also called \emph{double diffusive convection}, to occur in two forms: as layered double-diffusive (LDD) convection  characterized by convective layers and dynamic, turbulent interfaces where composition and temperature change drastically, and as oscillatory double diffusive (ODD) convection, where density perturbations oscillate around an equilibrium position.  Moreover, they have developed a prescription for the heat flux as a function of the gradients in density and temperature, which is crucial for determining the resulting temperature gradient ($\nabla_T:=d\ln T/d \ln P$) in the planet.  

At same heat flux, the superadiabaticity $\nabla_T-\nabla_{ad}$, where $\nabla_{ad}$ is the adiabatic temperature gradient, is enhanced in a semi-convective region compared to the case of full, overturning convection \citep{CB07,LC12}. A warmer-than-adiabatic interior as a result of semi-convection has been demonstrated to be able to prolong the cooling time of exoplanets \citep{CB07} and of Saturn \citep{LC13} by several Gyrs; it also allows one to add more heavy elements into the planet. In particular, \citet{LC12} (hereafter LC12) find that if LDD convection occurs throughout the interior of Jupiter, its heavy element content may be $2\times$ larger than derived from standard models.

In just a few years \emph{Juno} is expected to deliver new observational data on Jupiter.  Properties of interest (here the core mass, heavy element content, depth of zonal flows) can often not be measured directly but are inferred from model calculations that match the data.  
Now that an accurate He abundance measurement, comprehensive H/He demixing calculations, as well as semi-convective heat flux models all are at hand, we feel it is time to start to apply these three ingredients to begin to develop more advanced Jupiter models.
While this is a clear advance over previous work, we also caution that there is a forth leg to this ``chair'' that is missing in this work: 
we do not employ a theory for the formation, growth, and rain-out of He droplets here. 

With this fundamental caveat in mind, we apply in this paper a theory of double-diffusive convection as a result of assumed He rain, and investigate its 
effect on Jupiter's thermal evolution. We explore whether Jupiter's luminosity can be explained by the assumptions of Section \ref{sec:fundass}, which we think is a more self-consistent set of assumptions than conventional models rely on. This paper more aims at providing and discussing illustrative examples, rather than an evolved description of the physical processes inside the planet. We hope this paper will initiate the development  of the theoretical framework for the case of sedimentation in a giant planet. That task would vastly exceed the scope of this paper.

\paragraph*{Outline}
The computation of DD convection due to He rain is performed within a double iterative procedure.
In section \ref{sec:modelDD} we describe the inner loop, through which we ensure consistency between the temperature gradient and the heat flux. The theory of semi-convection (Sections \ref{sec:mathLDD}--\ref{sec:mathODD}) provides the superadiabaticity profile in the demixing region for given characteristic material parameters (Section \ref{sec:matprop}), 
given a heat flux profile (Section \ref{sec:cooling}), and a given He gradient profile (Section \ref{sec:HHeDemixing}).  
The \citet{Lorenzen09,Lorenzen11} H/He phase diagram that we use to compute the He abundance profile is described in Section \ref{sec:Lorenzen}. In Section \ref{sec:Yatm} we describe two modifications to it, and in Section \ref{sec:Yprofil} the outer loop that yields the consistency between the He profile and the temperature profile. 
Section \ref{sec:res1} contains our results for the application of the slightly modified H/He phase diagram (``modified-1''), and Section \ref{sec:res2} for the more severely (``modified-2'') H/He phase diagram . 
In Section \ref{sec:discuss} we discuss the results and suggest future steps. Section \ref{sec:summ} contains a summary.

\subsection{Fundamental assumptions}\label{sec:fundass}

Our method and results rely heavily on the following assumptions made in this work:\\
($i$) Jupiter's observed atmospheric He depletion is a result of He rain-out.\\
($ii$) The internal He abundance profile is dictated by the H/He phase diagram. \\
($iii$) The exchange of He droplets between  vertically moving eddies and the ambient fluid is negligible. 
This is the standard assumption in the Ledoux criterion.\\
($iv$) The internal temperature-pressure profile is not affected by the heavy elements.\\
($v$) Throughout the evolution, either LDD or ODD convection occurs in the demixing region.\\
($vi$) Jupiter's homogeneous deep interior below the rain zone remains adiabatic.

\subsection{Fundamental Caveats}

Conventional adiabatic models can well explain Jupiter's observed luminosity. Therefore, the additional
energy source implied by our assumptions $(i,ii)$ requires a compensating process for the total energy balance. 
The assumption ($v$) of semi-convection serves that purpose. However, one could in principle imagine a 
different scenario; for instance, core erosion \citep{Guillot03} could influence the energy balance as well. 
In fact, the applicability of the theory of semi-convection to the case of demixing and sedimentation has not
been proven yet.
This theory requires the diffusivity of solute to be less efficient than that of heat in order to maintain a 
composition gradient, while demixing and sedimentation imply an efficient albeit non-diffusive redistribution 
of solute. We nevertheless assume its applicability here on the grounds that a stabilizing compositional 
difference should exist between rising fluid elements and the surrounding medium. 
This is because an adiabatically evolving fluid element losing solute to condensation gains latent heat, stays warmer,
and hence can hold a higher equilibrium abundance of solute than the surrounding.
In other words, the non-diffusive nature of the condensation and rainfall leads to a reduction of the stabilizing composition gradient, although this reduction does not nullify it ($\beta>0$). As an approximation, 
we use $\beta=1$, where $0\leq \beta \leq 1$ is a scaling factor for the full predicted mean molecular weight gradient, and examine the results.

On the other hand, there is no known analogue. For instance, rain-forming water in the Earth is a minor 
constituent without stabilizing effect; demixing and sedimentation of Fe-Ni in the young Earth occured down 
to the centre without leaving behind a composition gradient in the mantle;   
and semi-convective regions in stars are often treated as zones of enhanced diffusion \citep{Langer85,DingLi14}, 
while here we assume diffusion to be negliglible compared to sedimentation. We return to these points in Sections 7.4--7.8.

\section{Material properties}\label{sec:matprop}

\begin{figure}
\includegraphics[width=0.45\textwidth]{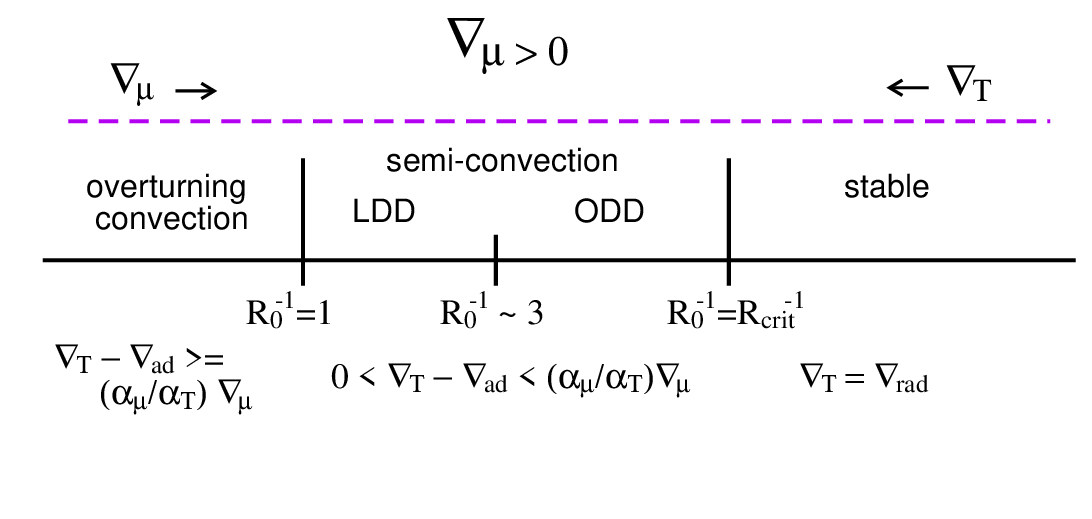}
\caption{\label{fig:theRegimes} Illustration of the four different regimes 
in the presence of a stabilizing composition gradient $\nabla_{\mu}$, 
which increases from left to right, while $\nabla_{T}$ increases from right to left.}
\end{figure}

The dimensionless \emph{Prantl number} 
\begin{equation}
	{\rm Pr} = \frac{\nu}{\kappa_T}
\end{equation} 
is the ratio of kinematic shear viscosity $\nu$ to thermal diffusivity $\kappa_T$, which have SI units of m$^2$/s.  In stars, Pr$\ll 1$, while in the water-rich interiors of Uranus and Neptune Pr~$>1$ might be possible  \citep{Soderlund13}. The dimensionless diffusivity ratio 
\begin{equation}
	\tau = D/\kappa_T
\end{equation}
measures the ionic particle diffusivity $D$ in relation to $\kappa_T$. In stars and gas giants, $\tau \ll 1$ because the ions are slower than the electrons and photons, and more heat is transported by electrons and photons than by the ions. For Jupiter, we neglect energy transport by photons. The thermal diffusivity $\kappa_T$ is related to the thermal conductivity $\lambda$ through 
\begin{equation}\label{eq:lamkap}
	\lambda = \rho\,c_P\,\kappa_T,
\end{equation}
where $\rho$ is mass density, $c_P$ specific heat, and $\lambda$ has SI units of WK/s. Both Pr and $\tau$ are important parameters because they define the transition between the double-diffusive (i.e.~semi-convective) and the stable regime. In fact, the transition occurs at the critical value 
\begin{equation}\label{eq:Rcrit}
	\Rcrit = \frac{1 + {\rm Pr}}{\tau + {\rm Pr}}\:,
\end{equation}
as initially discussed by \citet{Walin64} for the case of a stabilizing salt gradient in water heated from below,  
see Eq.~(13) therein.  
The \emph{inverse density-ratio} is defined as
\begin{equation}\label{eq:Rinv}
	\Rinv = \frac{\alpha_{\mu}}{\alpha_T}\frac{\nabla_{\mu}}{\nabla_T-\nabla_{ad}}
\end{equation}
(\citealp{Mirouh12}; LC12). $\Rinv$ includes the partial derivatives
\begin{equation}\label{eq:alphas}
	\alpha_{\mu}=\frac{\mu}{\rho}\frac{\partial \rho}{\partial \mu}|_{P,\,T}\quad,\quad
	\alpha_T = -\frac{T}{\rho}\, \frac{\partial \rho}{\partial T}|_{P,\,\mu}\:.
\end{equation}
While $\alpha_T$ can directly be calculated from the EOS, 
$\alpha_{\mu}$ is calculated using
\begin{equation}
	\frac{d\rho}{d\mu} = \frac{\rho^2}{\mu^2}\: 
	\left(\frac{\rho_{\rm H}^{-1} - \rho_{\rm He}^{-1}}{\mu_{\rm H}^{-1} - \mu_{\rm He}^{-1}}\right)\:.
\end{equation}
As a ratio of composition gradient ($\nabla_{\mu}$) to superadiabaticity ($\nabla_T-\nabla_{ad}$), $\Rinv$ is basically a density ratio between the differences in density due to different compositions and due to different temperatures that occur between a vertically moving parcel and its surrounding, respectively. The range $\Rinv\epsilon\: (1,\Rcrit]$ precisely defines the region of parameter space unstable to semi-convection. Ledoux instability implies $\Rinv\leq 1$, which marks the boundary between the overturning convective and the double diffusive regime. Thus, $\Rinv$ is the central quantity for determining whether a medium is in a state of double diffusive convection or not. An overview of these different regimes is given in Figure \ref{fig:theRegimes}.  In the stable regime, $\nabla_T=\nabla_{rad}$, where $\nabla_{rad}$ is the tempereture gradient needed to transport the heat by conduction and radiation. Section \ref{sec:modelDD} deals with  the method of deriving $(\nabla_{T}-\nabla_{ad})$, while Section \ref{sec:HHeDemixing} refers to $\nabla_{\mu}$.

Numerical values of the material properties in the 1--10~Mbar region along the Jupiter adiabat are given in Table~\ref{tab:matprops}. The values for $\lambda$, $\kappa_T$, $\nu$, and the particle diffusivities are taken from \citet{French12}, who computed the transport properties along the Jupiter adiabat using ab initio simulations. For the shear viscosity $\nu$ they found the dominant contribution to be the motions of the nuclei; other contributions are neglected in our applied values for $\nu$.

We also introduce a \emph{Rayleigh number} for LDD convection \citep{Wood13}
\begin{equation}
	{\rm Ra} = \frac{g\,\alpha\left(\frac{dT}{dr}-\frac{dT_{ad}}{dr}\right)l_H^4}{\kappa_T\,\nu}\:,
\end{equation}
where $g$ is gravity, $\alpha=\alpha_T/T$, and $l_H$ is the assumed height of the semi-convective layers.
Constraints on the value of $l_H$ are discussed throughout the paper.

\begin{table*}
 \begin{minipage}{\textwidth}
  \centering
  \caption{\label{tab:matprops}Material properties along the Jupiter adiabat. Data taken from \citet{French12}}
  \begin{tabular}{@{}ccccccccccc@{}}
  \hline
{$r$} &  {$T$} & {$P$} & {$\lambda$} & {$\kappa$} & {$\nu$} & {Pr} & {$D_H$} & {$\tau_{max}$} & {$\tau_{min}$} 
& {$R_{crit}^{-1}$}\\
 {$(R_J)$} & {(K)} & {(GPa)} & {(W/K/m)} & {(m$^2$/s)} & {(m$^2$/s)} & & {(m$^2$/s)} & {($=D_{\rm H}/\kappa$)} 
& {($=D_{\rm He}/\kappa$)}\\
\hline
0.196 & 18000 & 3410 & 1470 &  2.70e-05 & 0.266e-06 & 0.01  &    0.428e-06 & 0.0159 &   0.0101 &    40.4\\
0.350 & 16000 & 2460 & 1040 &  2.26e-05 & 0.282e-06 & 0.012 &    0.436e-06 & 0.0193 &   0.0111 &    32.6\\
0.478 & 14000 & 1640 &  721 &  1.89e-05 & 0.296e-06 & 0.0157 &   0.450e-06 & 0.0238 &   0.0134 &    26.7\\
0.584 & 12000 & 1030 &  465 &  1.50e-05 & 0.295e-06 & 0.0197 &   0.458e-06 & 0.0312 &   0.0165 &    20.4\\
0.680 & 10000 &  600 &  283 &  1.19e-05 & 0.313e-06 & 0.0263 &   0.468e-06 & 0.0393 &   0.0192 &    15.7\\
0.770 &  8000 &  300 &  153 &  8.56e-06 & 0.342e-06 & 0.04 &     0.481e-06 & 0.0562 &   0.0233 &    11.5\\
0.852 &  6000 &  120 &  59.6 & 4.99e-06 & 0.360e-06 & 0.072 &    0.471e-06 & 0.0944 &   0.0359 &     6.7\\
0.890 &  5000 &   64 &  20.2 & 2.16e-06 & 0.367e-06 & 0.17 &     0.369e-06 & 0.1708 &   0.0796 &     3.4\\
0.930 &  4500 &   23 &   3   & 3.55e-07 & 0.368e-06 & 1.036 &    0.274e-06 & 0.7718 &   0.73   &     1.2\\
\hline
\end{tabular}
\end{minipage}
\end{table*}


Our models are based on the SCvH EOS \citep{SCvH95}. Other EOS could be used as well if they provide the entropy 
for arbitrary H-He mixtures.

\section{Modeling Double Diffusive Convection}\label{sec:modelDD}

We aim to determine the local temperature gradient in a non-adiabatic planetary interior. For that purpose we make use of reference models (Section \ref{sec:refmodels}), and relations between the temperature gradient and the heat flux that can locally be transported along that gradient (Sections \ref{sec:Fcond}---\ref{sec:mathODD}). As the heat flux is constrained by the luminosity at the planet's photosphere, and as the energy loss of a planet ultimately leads to cooling and contraction, we also re-visit planetary cooling (Section \ref{sec:cooling}).

\subsection{Reference Jupiter models}\label{sec:refmodels}

To compute the demixing region in Jupiter we define two types of reference Jupiter models (two-layer and three-layer models). Furthermore, we make simplifying assumptions about its internal structure by putting all heavy elements of mass $M_Z$ as inferred from standard structure models into the core and assuming a pure H/He envelope of mean protosolar H/He ratio. In particular, we use reference models with a core mass of $28\:\ME$ or of $32\:\ME$, with $M_Z=28\:\ME$ being a typical value for SCvHi EOS based models \citep{SG04}, while a $32\:\ME$ core is found to best reproduce Jupiter's observed mean radius under the assumption of LDD convection.

Before demixing begins, Jupiter is described by a two-layer (2L) model with a rock core and one homogeneous adiabatic H/He envelope. For instance, our reference model '2L-ha-T180' for that case has a 1-bar temperature of 180 K and a $28\:\ME$ core mass. Our homogeneous, adiabatic two-layer reference model for the case that demixing does \emph{not occur in present} Jupiter of surface temperature \Tsurf=169 K is labelled '2L-ha-T169'. Finally, our quasi-homogeneous, adiabatic reference model for the case that demixing does occur but in the form of a sharp layer boundary at 1 Mbar between the depleted outer  and the enriched inner envelope is a three-layer model and labelled '3L-qha-T169'.

\subsection{Planetary cooling and luminosity profile}\label{sec:cooling}

The heat loss due to cooling and contraction of a planetary mass shell $dm$ at mass level $m$ per time 
interval $dt$ is given by $\delta Q(m) = T(m)\,ds(m)\:dm$, where $s(m)$ is the specific entropy of that mass shell, and $ds<0$ the change of the specific entropy during $dt>0$. This heat loss increases the planet's total luminosity by $dl_q = -\delta Q/dt$. With the specific energy of heat, $\delta q := \delta Q/dm = T\,ds$ the heat released from below the sphere of radius $r(m)$ is $l_q(m(r)) = -\int_0^{m(r)}dm' \frac{\delta q}{dt}$. Beside $l_q$, there can be further contributions to the luminosity of a giant planet or star such as sources from the decay of radioactive elements ($l_{\rm radio}$), nuclear reactions ($\epsilon_{\rm nucl}$; stars), or neutrino loss ($\epsilon_{\nu}$; as in neutron stars), so that in general
\begin{equation}\label{eq:dldm}
	\frac{dl}{dm} = -\frac{\delta q}{dt} + \frac{dl_{radio}}{dm} + \epsilon_{\rm nucl} - \epsilon_{\nu} + \ldots\:.
\end{equation}
For the majority of stars, the first two terms can be neglected, and the local luminosity be computed, in parallel with the temperature and compositional profiles, as an integral over the nuclear reaction rates. For giant planets however, only the first two terms in Eq.~\ref{eq:dldm} play a role, so that the luminosity becomes
\begin{equation}\label{eq:lm}
	l(m(r)) = -\int_0^{m(r)}dm'\, \left[\frac{\delta q}{dt} - \frac{dl_{radio}}{dm'}\right]\:.
\end{equation}
To obtain Jupiter's current luminosity profile, we evolve the planets' internal structure down from the state before demixing began. The time interval $dt$ between two subsequent internal states appears in Equation \ref{eq:lm} as a scaling factor. As in \citet{N+12} we use $dt$ to adjust the known intrinsic luminosity $L_{int}$,
\begin{equation}\label{eq:dt}
dt = \frac{\int_0^{M_p}dm'\, Tds}{L_{\rm int} - L_{\rm radio}}.
\end{equation} 
where $L_{\rm int}=L_{\rm eff} - L_{\rm eq}$ and $L_{\rm eff}$ is either the observed luminosity at present time ($F_{\rm eff}=L_{\rm eff}/4\pi R_p^2=13.6$ W/m$^2$ for Jupiter), or the predicted one of a model atmosphere, as required for instance for the evolving planet at earlier times. $L_{\rm eq}$ describes the incident flux ($F_{\rm eq}=8.2$ W/m$^2$ for Jupiter) that is derived from the stellar luminosity, orbital distance, and Bond albedo.

To conclude, for given temperature and entropy profiles, Equations (\ref{eq:lm}) and (\ref{eq:dt}) provide us with the internal luminosity and heat flux profiles, $F(m)=l(m)/4\pi r^2(m)$.

\subsection{Thermal evolution}\label{sec:methEvol}

To compute Jupiter's thermal evolution with H/He phase separation and DD convection we generate interior models for different surface temperature down to {\Tsurf=169} K. These models provide the internal profiles of temperature and entropy, which are needed to compute the inhomogeneous evolution, i.e.~the evolution when the composition changes with depth. Note that for homogeneous evolution it suffices to know the entropy only up to a constant offset value which may depend on composition, as that offset value cancels out when taking the difference $T\,ds$. 

For sufficiently high surface temperatures, interior temperature are too high for demixing to occur. Thus we represent Jupiter's evolution prior to the onset of demixing by a series of adiabatic, homogeneous 2L models with a rock core.
To compare the evolution with and without He rain we also expand that series down to \Tsurf $= 169$ K.


The cooling of the planet is then computed as described in \citet{N+12}, but here we neglect angular momentum conservation. For the outer boundary condition we use either the \citet{Graboske75} model atmosphere grid, or the non-grey atmosphere model of \citet{Fortney+11}, which these authors found to yield a $\sim 500$ Myr longer cooling time for Jupiter.

\subsection{Conductive heat transport}\label{sec:Fcond}

The local temperature gradient $dT/dr$ depends on the processes through which the heat is transported. Possible heat transport mechanisms in giant planets are  radiation, conduction, ODD convection, LDD convection, and overturning convection.  The relation between heat flux and temperature gradient in the one-dimensional conductive case reads
\begin{equation}\label{eq:Fcond}
	F_{cond} = -\lambda \: dT/dr\:.
\end{equation}
Equation (\ref{eq:Fcond}) yields the heat flux that is transported by conduction along a known temperature gradient. In the case of predominantly conductive heat transport we could invert Eq.~(\ref{eq:Fcond}) to obtain the temperature gradient. However, conductive heat transport, and also radiative heat transport, is usually inefficient in giant planets so that the temperature gradient needs to be determined by other means.

\subsection{Relation between heat flux and temperature gradient in LDD convection}\label{sec:mathLDD}

We derive an expression for the relation between the heat flux in case of LDD convection, $\FLDD$, and the temperature gradient, following closely the description of \citet{Wood13}, their equations (16)--(18).
In LDD convection, convective layers are separated by interfaces, and thin adjacent boundary layers, of strongly varying temperature and composition gradients. Thus the temperature gradient is not continuous on the scale of individual ''steps'' in the staircase. However, it is possible to consider an average temperature gradient, if taken over at least one subsequent pair of layers plus interfaces, as they are found to occur in computer simulations \citep{Mirouh12,Wood13}. It is this average temperature gradient we are interested in.

In a convective medium where heat is transported by both conduction and turbulent motions the total heat flux $F$ is given by $F = F_{cond} + F_{conv}$. Usually, a mixing length theory (MLT) based expression for $F_{conv}$ is used. In case of LDD convection, the turbulent heat flux reduces to $\FLDD$. Thus we have
\begin{equation}\label{eq:FT1} 
	F = F_{cond} + \FLDD\:.
\end{equation}
Using the notation of \cite{Wood13}, $F_{cond}=-\rho\,c_P\kappa_T(dT/dr)$ where $\rho$ is the local density, and 
\begin{equation}\label{eq:FLDD1}
	\FLDD = \rho\,c_P\,\kappa_T\,(1 - {\rm Nu}_T)\left(\frac{dT}{dr} - \frac{dT_{ad}}{dr} \right)\:,
\end{equation}
where Nu$_T$ is the thermal Nusselt number, whose expression is discussed below.
Inserting Eqs.~(\ref{eq:FLDD1}, \ref{eq:Fcond}) into Eq.~(\ref{eq:FT1}) and using Eq.~(\ref{eq:lamkap}) we obtain
\begin{equation}\label{eq:FT2}
	F = -\lambda \frac{dT}{dr} - \lambda\,({\rm Nu}_T-1)\left(\frac{dT}{dr} - \frac{dT_{ad}}{dr} \right)\:.
\end{equation}
Next we seek to express Nu$_T$ in terms of parameters that can be evaluated, namely of those introduced in 
Section \ref{sec:matprop}. \citet{Wood13} found that the expression  
\begin{equation}\label{eq:NuTansatz}
	{\rm Nu}_T - 1  = f_T(\Rinv; \tau) \: {\rm Ra}^a\: {\rm Pr}^b
\end{equation} 
with $a=0.34\pm 0.01$ and $b=0.34\pm 0.03$, provided a reasonable fit to their numerical experiments.  The function
$f_T(\Rinv;\tau)$ remained poorly constrained but was found to take values between 0 and 0.2 for $\Rinv=1.1$--2 and $\tau=0.01$--0.3, with $f_T$ decreasing with $\Rinv$, but fairly independent on $\tau$. Various functional formula for $f_T$ will be tested. Inserting Eq.~(\ref{eq:NuTansatz}) 
into (\ref{eq:FT2}) gives
\begin{equation}\label{eq:FT3}
	F = -\lambda \frac{dT}{dr} 
	- \lambda\,f_T(\Rinv,\tau)\:({\rm Ra}\,{\rm Pr})^{1/3}\left(\frac{dT}{dr} - \frac{dT_{ad}}{dr} \right)\:. 
\end{equation}
With the product ${\rm Ra}_{\star}=:$ $\rm Ra\, Pr=$ $g\alpha l_H^4$ $\times$ $\left(dT/dr - dT_{ad}/dr\right)/\kappa_T^2$ Equation (\ref{eq:FT3}) then becomes
\begin{equation}\label{eq:FT4}
	F = -\lambda \frac{dT}{dr} - \lambda\,f_T\left(\frac{g\,\alpha\,l_H^4}{\kappa_T^2}\right)^{1/3}
	\left(\frac{dT}{dr} - \frac{dT_{ad}}{dr} \right)^{4/3}
\end{equation}
which can be evaluated and solved for the temperature gradient $dT/dr$ numerically. 
Note that the first term in Equation (\ref{eq:FT4}) is $F_{cond}$ and the second one is $F_{\rm LDD}$.
Equation~\ref{eq:FT4} is equivalent to Equation (7) in LC12.

\subsection{Relation between heat flux and temperature gradient in ODD convection}\label{sec:mathODD}

The heat flux in the case of ODD convection can be expressed in terms of the Nusselt number in the same way as in case of LDD convection (Equation \ref{eq:FLDD1}). For Nu$_T$ we adopt the fit to the simulation data of \citet{Mirouh12},
\begin{equation}\label{eq:NuT_ODD}
	Nu_T-1 \simeq 0.75 \left(\frac{\rm Pr}{\tau}\right)^{\!0.25\pm 0.15} \!\!\frac{1-\tau}{\Rinv-1}
	\left(1-\frac{\Rinv-1}{\Rcrit-1}\right)\:.
\end{equation}
The procedure then is the same as described in Section \ref{sec:mathLDD}, only that Eq.~(\ref{eq:NuT_ODD}) instead of Eq.~(\ref{eq:NuTansatz}) is inserted into Eq.~(\ref{eq:FT2}).
The resulting values of $\Rinv$ can then be used as a self-consistency check for our fundamental assumption $(v)$.

\subsection{Academic exercises}\label{sec:toys}

In order to understand the behaviour of the possible superadiabaticity in Jupiter as a function of layer height and composition gradient, we investigate toy models first. In Section \ref{sec:fnconst} we assume $f_T$ to be constant. In Section \ref{sec:fT} we will account for the dependence of $f_T$ on $\Rinv$ explicitly.

\subsubsection{Constant $f_T$ values}\label{sec:fnconst}

\begin{figure}
\includegraphics[width=0.45\textwidth]{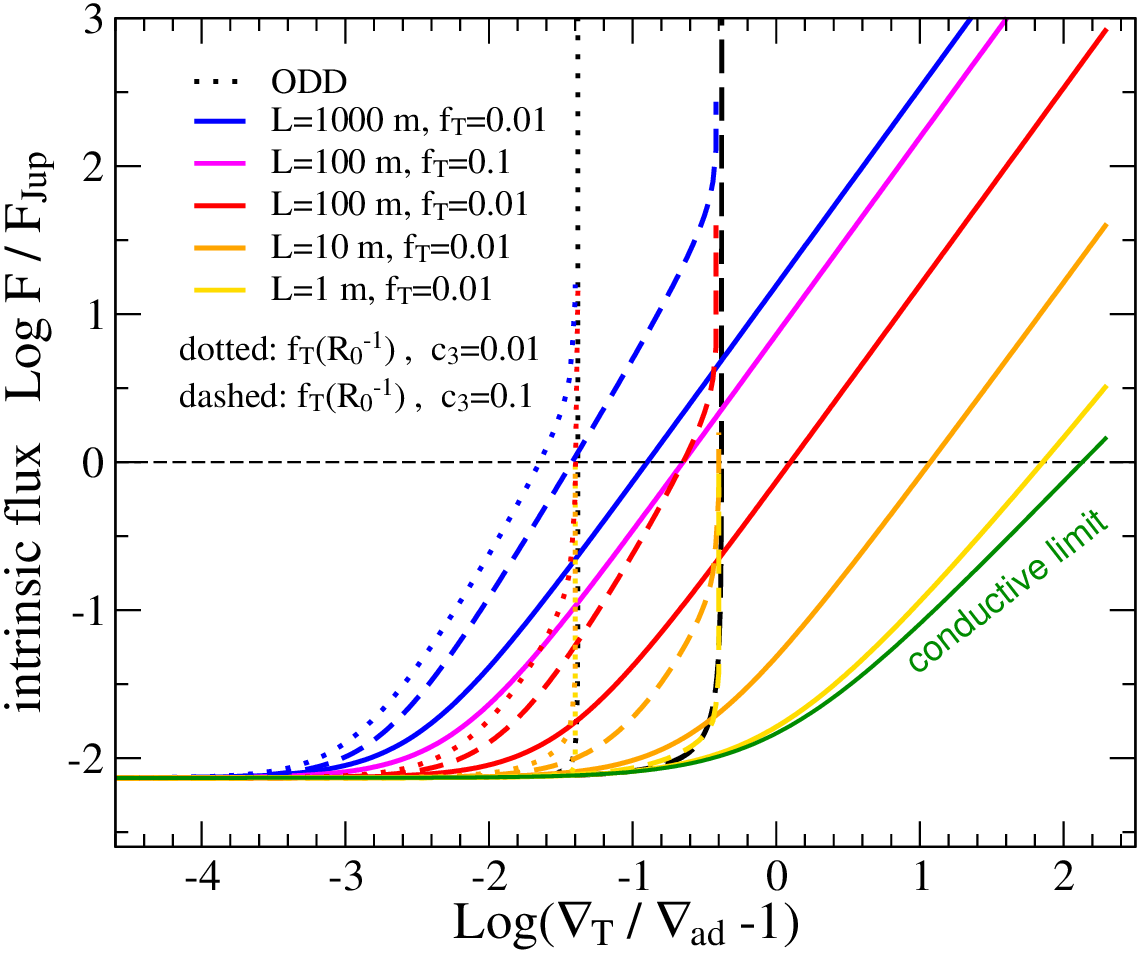}
\caption{\label{fig:flux_superadA}
Illustration of Eq.~\ref{eq:FT4} and of Eqs.~\ref{eq:FT2} with \ref{eq:NuT_ODD}: intrinsic heat flux scaled by Jupiter's surface heat flux as a function of superadiabaticity, for different layer heights as labelled 
and for different assumptions on $f_T$: 
\emph{solid lines}: constant $f_T$ values of 0.1 (\emph{magenta only}) and 0.01; \emph{dashed and dotted lines}: including the functional dependence of $f_T$ on $\Rinv$ by assuming $c_3=0.1$ (\emph{dashed}), or $c_3=0.01$ (\emph{dotted}), see Eq.~(\ref{eq:c3}). The \emph{dark green} line shows the conductive limit.
This figure shows that Eq.~(\ref{eq:FT2}) can easily be inverted numerically to find the temperature gradient 
for given values of $F$, $l_H$, and $f_T$. Graphically, this temperature gradient occurs where the respective 
\emph{coloured} curve (LDD convection) or the respective \emph{black} curve (ODD convection) crosses the  \emph{vertical black line}.
The relations are displayed for a wider range in $\Rinv$ and $l_H$ values than physically allowed (see text).
}
\end{figure}

Figure \ref{fig:flux_superadA} displays the relation (\ref{eq:FT4}) between the total flux $F$, scaled by Jupiter's intrinsic flux at the surface, $F_{\rm Jup}=5.44$ W/m$^2$, and the relative superadiabaticity $(	\frac{dT/dr - dT_{ad}/dr}{dT_{ad}/dr} = \nabla_T/\nabla_{ad} -1$) using values of the material parameters of Table \ref{tab:matprops} that are typical for the 1--2 Mbar region in Jupiter, where H/He demixing is supposed to occur. 
\emph{Solid lines} are for constant $f_T$-values. \emph{Other lines} will be explained in Section \ref{sec:fT}.

According to Figure (\ref{fig:flux_superadA}), the flux $F_{\rm LDD}$ increases with $\nabla_T$. For 
small $\nabla_T/\nabla_{ad}-1\ll 10^{-3}$, $F_{\rm LDD}\ll F_{cond}$ for all layer heights so that 
$F \approx F_{cond}\approx 10^{-2}F_{\rm Jup}$. With increasing relative superadiabaticity,  
$F\approx F_{\rm LDD}\sim f_T\l_H^{4/3}$: the smaller the layer height, and the smaller $f_T$, 
the lower the heat flux. Through layer heights below 1~m the heat flux is as inefficient as 
conductive heat transport and would require high relative superadiabaticities of 10--100 to allow 
Jupiter's observed heat flux be transported.
On the other hand, layer heights larger than 1000~m would imply $\nabla_T \approx \nabla_{ad}$ and thus are expected to have little effect on Jupiter's temperature profile compared to adiabatic standard models.

In principle, the range in possible layer heights is further restricted by the requirement that layers 
can form at all, which is seen to occur in simulations not below a minimum length scale of about $5/3\times 20\times$ the \emph{instability length scale parameter} $d$ \citep{Wood13},
\begin{equation}\label{eq:dmin}
	d = \left(\frac{\kappa_T\:\nu}{\alpha\:g |dT/dr - dT_{ad}/dr|}\right)^{1/4}\:.
\end{equation}
\citet{Wood13} point out that the value of $L_H$ is just slightly larger than the wavelength of the fastest growing linear mode ($20d$), which is the most important one because it rapidly dominates the dynamics of the system due to the exponential amplification of the initial state, at least within linear instability analysis. With $d\approx 5$--50 cm, this lower limit on $l_H$ of 1.5--15 m agrees well with the minimum layer height of 1 m found in Figure \ref{fig:flux_superadA}. For illustration however, we present here the relations for a wider range of layer heights, and also of $\Rinv$ values, than actually allowed.

\subsubsection{$f_T$ as a function of superadiabaticity for constant composition gradient}
\label{sec:fT}

\begin{figure}
\includegraphics[width=0.45\textwidth]{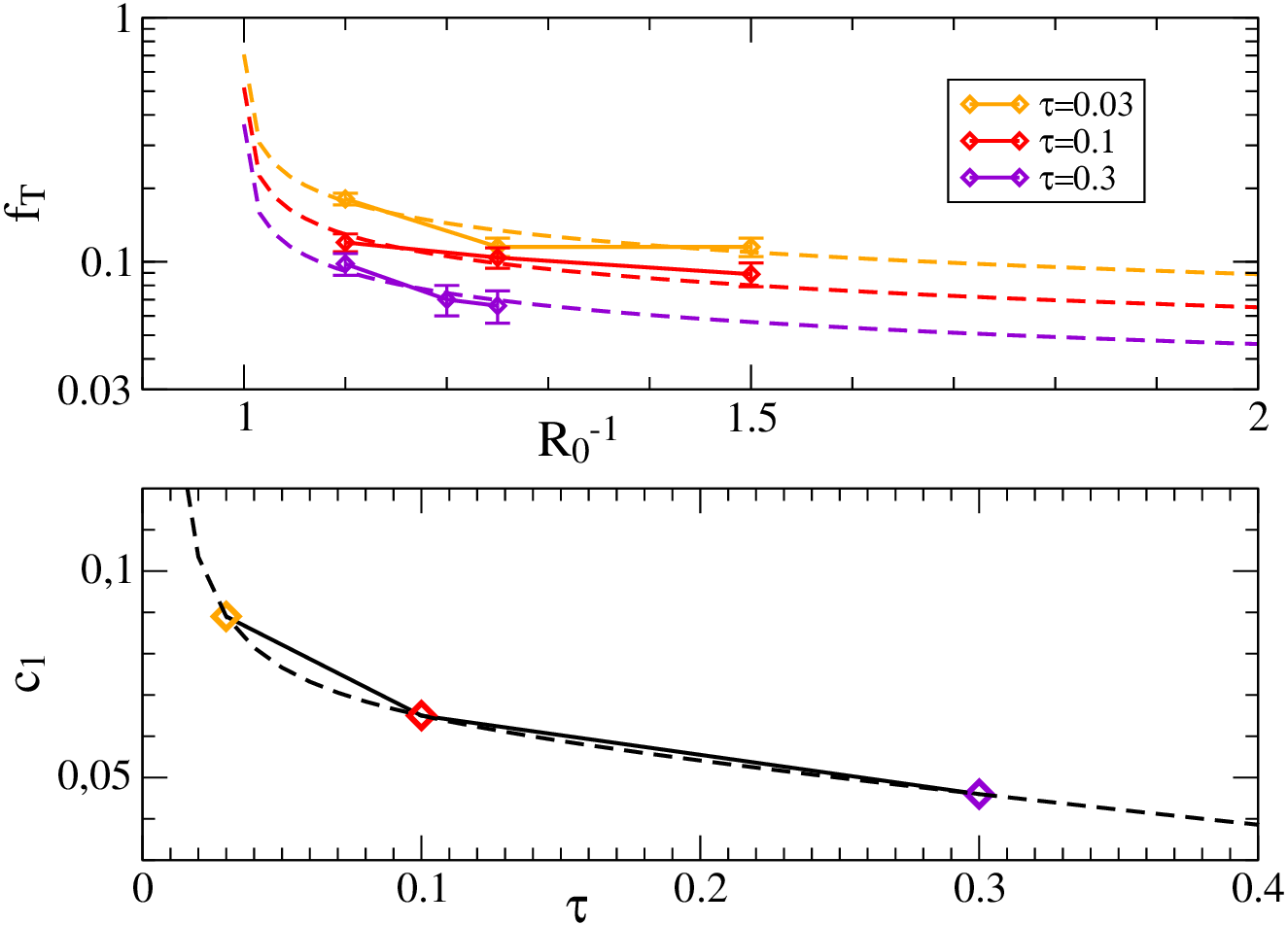}
\caption{\label{fig:fTfit}Fit function $f_T(\Rinv,\tau)$
Upper panel: Function $f_T(\Rinv,\tau)$ (\emph{dashed}) and the data points that $f_T$ is designed to fit
from Figure 6 in \citet{Wood13} (symbols); lower panel: fit coefficient $c_1(\tau)$, see Eq.~(\ref{eq:fTfit}). }
\end{figure}

We replace the formerly constant $f_T$ values by a function $f_T(\Rinv,\tau)$ which is obtained by fitting
the simulation data of \cite{Wood13}, their Figure 6, within $\tau=$ 0.03--0.3 and $\Rinv=$ 1.1--1.5.
We choose a functional form that guaranties $f_T\to 0$ for $\Rinv\to\infty$ and $f_T\to 1$ for 
$\Rinv\to 1$. $\Rinv$ values lower than 1 are not of interest because the medium would then be
in the overturning convection state, while for high $\Rinv$ values semi-convection ceases in favor of a diffusive heat transport $(f_T=0)$.  Thus we use the function 
\begin{equation}\label{eq:fTfit}
	f_T(\Rinv,\tau):=\frac{c_1(\tau)}{\left[\Rinv-(1-\varepsilon)\right]^{c_2}}
\end{equation}
and adjust $c_1(\tau)$ and $c_2$ to match the simulation data.  The small parameter $\varepsilon=10^{-3}$ 
ensures that $f_T(\Rinv=1)$ is well behaved and close to 1 as in the usual MLT, although even for $f_T=1$ Equation \ref{eq:FT4} would not exactly describe the MLT case because of the different exponents of Ra$_{\star}$. We find $c_2=0.3$ and $c_1(\tau)=c_{11} + c_{12}\tau + c_{13}\tau^{-1}$ with $c_{11}=0.06348$, $c_{12}=-0.06746$, $c_{13}=0.0008262$. Our fit function $f_T$ is shown in Figure \ref{fig:fTfit}. In the demixing region, $f_T$ typically decreases weakly by a factor of 2 and adopts values in the extrapolated regime at $\Rinv> 1.5$. As $F_{\rm LDD}$ scales with $f_T\times l_H^{4/3}$, any uncertainty in $f_T$ can be expressed as an uncertainty in $l_H$ and thus should not affect the  resulting possible range of superadiabaticities.

By using our fit-formula $f_T(\Rinv,\tau)$, we already include the dependence of $f_T$ on the composition gradient.
For our academic exercise, we simplify this dependence by setting
\begin{equation}\label{eq:c3}
	\Rinv = \frac{c_3}{\nabla_T-\nabla_{ad}}\:,
\end{equation}
with a constant toy composition gradient $c_3$ for which we assume $c_3=0.01$ and $c_3=0.1$, respectively, in close agreement to the values that we calculate for the demixing region in Jupiter. Here we investigate the effect of a given constant composition gradient on the intrinsic flux (Eq.~\ref{eq:FT4}) and on the possible superadiabaticity. 

\begin{figure}
\includegraphics[width=0.40\textwidth]{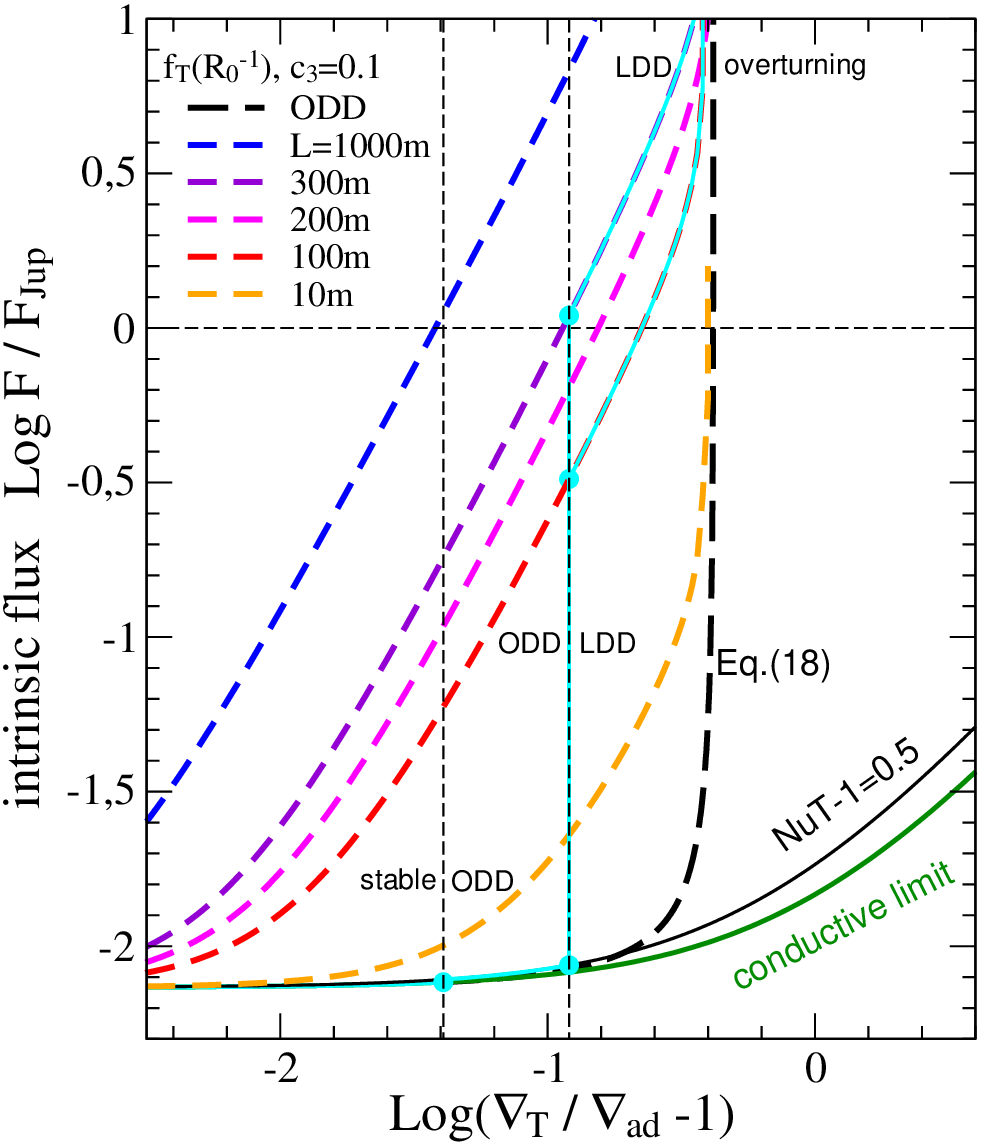}
\caption{\label{fig:flux_superadB}
Similar to Fig.~\ref{fig:flux_superadA} but for $c_1=0.1$ only. The \emph{thin solid black} curve is for ODD convection assuming NuT$-1=0.5$. The \emph{left vertical line} marks the transition between the stable and the ODD regime ($\Rinv=\Rcrit$) while the \emph{right vertical line} marks the transition between ODD and LDD convection ($\Rinv=3$). Thus, only the \emph{cyan curve} displays the physically allowed relation, which depends on the assumed layer height in the LDD regime.}
\end{figure}

We go back to Figure \ref{fig:flux_superadA} to examine the \emph{dashed} and \emph{dotted} curves therein. In addition, Figure \ref{fig:flux_superadB} shows a zoom-in for $c_3=0.1$.
First, the dashed and dotted curves have a steeper slope than those for constant $f_T$ values, 
probably because of $f_T \sim (\Rinv)^{-0.3}$, and thus $F_{\rm LDD}\sim (\nabla_T-\nabla_{ad})^{5/3}$ 
instead of 4/3 (compare Eq.~\ref{eq:FT4}). Second, the smaller the composition gradient (smaller $c_3$ value), the higher the flux at fixed superadiabaticity. 

Third, the smaller $c_3$ (dashed $\to$ dotted) and the larger the superadiabaticity (left $\to$ right),  the smaller becomes the $\Rinv$ value. Eventually $\Rinv\to1$ happens. Because of our functional choice for $f_T$, which prohibits $\Rinv<1$ by letting $f_T$ rise to infinity, the slope of $F$ then tends to infinity. This behaviour tells us that the medium wants to transition to overturning convection regime. We recover here the well-known fact that for a given composition gradient, there is an upper limit to the superadiabaticity in LDD convection. The lower $c_3$, the lower the maximum possible superadiabaticity. 
For small layer heights (here for $l_H<100$ m), this upper limit on the superadiabaticity implies that the desired flux can not be transported by semi-convection, which implies a lower limit on $l_H$.

Fourth, for $l_H\gtrsim 100$ m and $F\approx F_{\rm Jup}$, Figure \ref{fig:flux_superadA} shows 
$F_{cond}\ll F\approx F_{\rm LDD}$. In that case, $l_H^{4}\sim (\nabla_T-\nabla_{ad})^{-5}$. With increasing
layer height at a given composition gradient, the needed superadiabaticity might eventually become so small that $\Rinv$ gets larger than $\Rcrit$, which contradicts the assumption of LDD convection. In particular, $\Rinv$ close to $\Rcrit$ is usually associated with ODD convection (\citealp{Mirouh12}; LC12). We therefore include ODD convection in our considerations.

Fifth, Figure \ref{fig:flux_superadB} indicates that layer heights above 1 km may yield $\Rinv > \Rcrit$ if $F=F_{\rm Jup}$ is to be transported while ODD convection appears to be too ineffcient to transport $F_{\rm Jup}$. The \emph{cyan curve} highlights the heat flux--superadiabaticity relation for which the $\Rinv$ value would be consistent with the respective regime (stable, ODD, or LDD). In our toy models, a narrow range of $l_H\approx 100$--300 m emerges for which F$_{\rm Jup}$ can be transported. 

From here on, we can methodically proceed in two different directions: we could use only those relations like the \emph{cyan curve} in Figure \ref{fig:flux_superadB} that a guarentees $\Rinv$ values consistent with the assumed regime (stable, LDD, ODD). That approach narrows down the $l_H$ value before any self-consistent, converged  model is found. Here we decide to trod a different way: we first construct models with He-rain for a wide range of $l_H$ values, and then ask whether the fully converged model satisfies the consistency criterion for $\Rinv$. We think that this approach makes it easier to understand the behaviour of the solutions, as they smoothly transition from already explored territory, where the effect of He rain on the planet's thermal evolution is dominated by the gravitational energy release  \citep{Hubbard99,FH03}, to new territory, where we will see the effect to be dominated by the internal temperature profile.

\subsection{Consistency between F(T) and T(F): the inner Loop}\label{sec:innerloop}

In Sections \ref{sec:mathLDD} and \ref{sec:mathODD} we have presented the relation between heat flux and temperature gradient in LDD and ODD convection, respectively (for a given $\nabla_{\mu}$). One can walk the trail in either direction and use these relations to derive the heat flux from a given temperature gradient  (and composition gradient), or the temperature gradient for a given heat flux (and $\nabla_{\mu}$), depending on what is known a priori.
Both the heat flux and the temperature profiles are a priori unknown in Jupiter, unless the interior is adiabatic (ignoring here any uncertainty due to the EOS).

We showed in Section \ref{sec:cooling} how we can determine the heat flux profile for given profiles of temperature and entropy. We compute a first guess on $F(m)$ by using the reference model 3L-qha-T169 and Equation \ref{eq:lm}. Then, an iterative procedure is performed that iterates between the computation of the temperature gradient from the heat flux profile (Equation \ref{eq:FT2}) and the computation of the heat flux profile from the temperature and entropy\footnote{In detail, $T(P)$ is computed by a Fourth-order Runge-Kutta-Integration of $\nabla_T$, and then $s(T,P)$ is derived from the EOS.} profiles (Equation \ref{eq:lm}), until a converged solution is obtained. For this \emph{inner loop} the helium abundance profile is kept constant\footnote{In detail, $Y$ is kept constant as a function of mass $m$ to ensure helium mass conservation. Intermediate planet models are computed to ensure additional consistency between $m$ and $P$, as it is $Y(P)$ that is provided by the H/He phase diagram, not $Y(m)$.}.

\section{Modeling the helium abundance profile}\label{sec:HHeDemixing}

Helium is predicted to demix from hydrogen at high pressures ($\sim 1$ Mbar) and sufficiently low temperatures in regions where hydrogen undergoes pressure ionization from the molecular to the metallic state, where He is still non-metallic \citep{Salpeter73,Lorenzen11}. Demixing is also seen at lower pressures where hydrogen is in the molecular phase, both in numerical simulations \citep{Morales13} and in laboratory experiments \citep{Loubeyre85}, although at much lower temperatures. It is even predicted to occur in fully ionized H-He mixtures up to 200 Mbar \citep{Stevenson75}.

Published H-He phase diagrams largely agree in predicting demixing of H and He at temperatures of several 1000 K and pressures of a few Mbar, the typical $T$--$P$ space of evolved giant planets like Jupiter and Saturn. However, the predictions of the slope and the locations of the phase boundaries for the demixing temperature as a function of pressure and helium abundance have changed considerably over time, and with them the predictions for the presence and extension of demixing zones in Jupiter and Saturn. Results are diverse, and include demixing in both planets within at least 5--20 Mbar 
\citep{Klepeis91}, no demixing in either of them  \citep{Pfaffenzeller96}, demixing in Saturn down to the core with no demixing in Jupiter \citep{Morales13}, and demixing in both planets at depths below 1 Mbar \citep{Lorenzen11}.  The more modern calculations (Morales et al., Lorenzen et al.) agree much better with each other than the older ones.

We here apply the H-He phase diagram of \citet{Lorenzen09,Lorenzen11} for three reasons: it provides a very dense grid of  demixing temperatures $T_{\rm dmx}(P,x_{\rm He})$, that is the maximum temperature below which H/He phase separation occurs, as a function of He abundances $x_{\rm He}$ for relevant pressures $P$; it does predict demixing in Jupiter and allows for the computation of the helium abundance profile and atmospheric depletion; finally, it is based on state-of-the art first-principles simulations using classical molecular dynamics simulations for the ionic subsystem and density functional theory for the electronic subsystem (DFT-MD simulations), a method that has repeatedly yielded data in remarkably good agreement with experiments, such as for pure hydrogen (\citealp[see, e.g.,][]{Becker13}). 

Independently, \citet{Morales09,Morales13} also computed the H-He phase diagram by using similar ab initio simulation methods to derive the Gibbs free energy, which yields the energetic preference of mixing or demixing. The main two differences between both groups lie in (i) the determination of the entropy of mixing and (ii) in the functional form used to fit the the enthalpy of mixing $\Delta H$ as a function of $x_{\rm He}$
for given $P,T$ values. While \citet{Lorenzen09,Lorenzen11} neglect non-ideal contributions to the entropy of mixing but capture the asymmetric shape of $\Delta H(x_{\rm He})$ through a high-order expansion, \cite{Morales09,Morales13} include the non-ideal entropy of mixing  but approximate $\Delta H(x_{\rm He})$ by a quadratic fit only, which appears a reasonable match to their sparse data sample but not to the fine Lorenzen data grid.
Both approximations affect the double tangent construction of the Gibbs free energy $\Delta G(\Delta H, S_{\rm mix})$, from which the energetic preference for demixing and the corresponding equilibrium compositions are derived. 
While experimental efforts are under way \citep{Soubiran13}, in the absence of experimental constraints on H/He demixing under planetary interior conditions we consider the deviations in $T_{\rm dmx}(x_{\rm He}; P)$ between the two theory groups as an indication for the real uncertainty; it amounts to $\sim 500$ K at 4 Mbar and even 1000 K at 1 Mbar.

\subsection{The Lorenzen et al H/He demixing diagram}\label{sec:Lorenzen}

\begin{figure}
\includegraphics[width=0.45\textwidth]{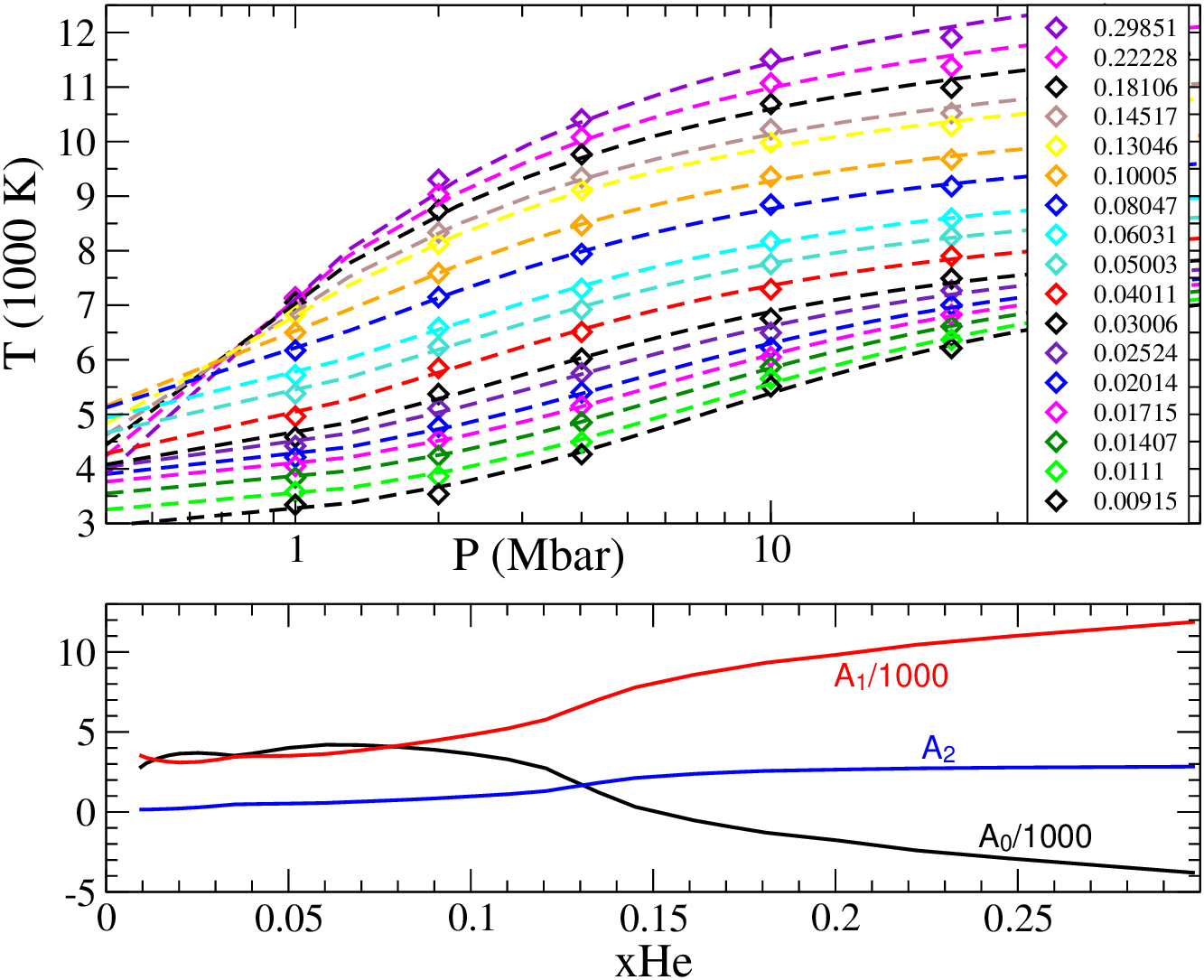}
\caption{\label{fig:Tdmx}
Illustration of the 3-parameter fit to the \citet{Lorenzen09,Lorenzen11} H/He phase diagram data \emph{(diamonds)}
in $T(P;x_{\rm He})$ for constant $x_{\rm He}$ (upper panel) and the three fit-coefficients of Eq.~(\ref{eq:fitTdmx})
as a function of $x_{\rm He}$ (lower panel).}
\end{figure}

We here describe our semi-analytical fit to the \citet{Lorenzen09,Lorenzen11} data for $T_{\rm dmx}(x_{\rm He}; P)$. The published data span a grid of pressures \{1, 2, 4, 10, 24 Mbar\}, and a very dense grid in helium abundance ranging from pure hydrogen ($x_{\rm He}=0$) to pure helium ($x_{\rm He}=1$). Applying the raw data with simple interpolation to Jupiter, we find that the adiabat intersects with the demixing region between 1 and about 3.5 Mbar; thus all information on the helium abundance profile $Y(P)$ is based on only 3--4 simulated pressure grid points. Since we need smooth gradients $\nabla_{\mu}$ and $\Rinv$, we are forced to develop a semi-analytical fit to the data in terms of $T_{\rm dmx}(P; x_{\rm He})$ as shown in Figure \ref{fig:Tdmx}. Fortunately, the highest helium abundances found to occur in present Jupiter are $x_{\rm He}<0.3$ so that we can ignore the high-$x_{\rm He}$ part of the demixing diagram when fitting the data. First we choose a representative subset in $x_{\rm He}$ and display $T_{\rm dmx}(P; x_{\rm He})$ in Figure \ref{fig:Tdmx}, upper panel. For each of these $x_{\rm He}$-values we then fit $T_{\rm dmx}(P; x_{\rm He})$ by the fit formula
\begin{multline}\label{eq:fitTdmx}
	T_{\rm dmx}(P; x_{\rm He}) = A_0(x_{\rm He})  +\\
	A_1(x_{\rm He})\times\arctan\left(\log(1 + A_2(x_{\rm He})\, P)\,\right).
\end{multline}
Although the $\arctan$-function is non-unique (it maps onto itself under variation of the argument),
we found a reasonable behaviour of the coefficients $A_i(x_{\rm He})$, see Figure \ref{fig:Tdmx}. Indeed, none of the other functional forms we tried yielded a better behaved $Y(P)$ profile. Other fit formulas we tried might show an almost indistinguishable behaviour in $T_{dmx}(P)$ but would yield minima or maxima in $Y(P)$ instead of a monotonic behaviour. This implies a strong sensitivity of the resulting He abundance profile in the planet on the functional form used but also lends confidence to the one chosen.

\subsection{Atmospheric helium depletion}\label{sec:Yatm}

\subsubsection{Assumptions}

The H/He demixing diagram allows for the calculation of the helium abundance over a  planet's entire internal pressure range.  As in \citet{SS77b} we here assume that if demixing occurs, He droplets will form and sink to a depth
where demixing is no longer predicted to occur, or the core boundary is reached. 
More specifically, we assume that He droplets will sink as long as $T(P,x_{\rm He}(P)) < T_{\rm dmx}(P,x_{\rm He}(P))$. Demixing terminates when the phase boundary between mixed and demixed state is reached, i.e.~if $T(P,x_{\rm He}) = T_{\rm dmx}(P,x_{\rm He})$. This describes the equilibrium state that we require our planetary $P$--$T$--$x_{\rm He}$ profiles to achieve. 

To  obtain the atmospheric helium mass fraction,  $\Yatm$ due to He rain we compute an initial H/He adiabat for Jupiter's known 1-bar temperature and protosolar H/He ratio, while heavy elements  are neglected, as their distribution in response to He rain is unknown and its investigation beyond the scope of this paper. Since our initial adiabat intersects with the demixing curve ($T(P,x_{\rm He}^{\rm (proto)}) < T_{\rm dmx}(P,x_{\rm He}^{\rm (proto)})$) over some pressure range, we lower the $x_{\rm He}$ value of the adiabat and iterate between the adiabat and demixing curve until the $T$--$P$ profile of the adiabat just touches the demixing curve. This provides us with a unique, converged value $x_{\rm He}=x_{\rm He}^{\rm (A)}$ for a given 1-bar surface temperature. Abundances $x_{\rm He} < x_{\rm He}^{(\rm A)}$ would lead to no crossover between the adiabat and demixing curve, while higher He abundances to an intersection. This behaviour is a result of the strong decrease in $T_{\rm dmx}$ toward lower $x_{\rm He}$ values in the relevant $x_{\rm He}$--range (Figure \ref{fig:Tdmx}), which more than compensates the cooling of the adiabat with lower $x_{\rm He}$ values. 
For the unmodified \citet{Lorenzen09,Lorenzen11} data, this touch-point (the equilibrium state) occurs at $P_0=1$ Mbar. 
As in \citet{SS77b} we assume that all planetary material, atop the onset pressure $P_0$ for demixing will be mixed over time into the demixing region through convection. We here make the assumption of \emph{instantaneous} sedimentation, meaning that the background profile follows the phase diagram as a result of assumed rapid He droplet formation and assumed rapid sinking to a level where they dissolve, before convection could redistribute the droplets upward (but see also Section \ref{sec:Rparam} and \ref{sec:Earth}). Therefore, the excess He from outer envelope material that gets mixed into the He rain zone through convection will sink down. Over time, the He abundance in the planetary atmosphere and in the entire envelope down to $P_0$ decreases to the value of $x_{\rm He}^{\rm (A)}$. Because of Jupiter's  short convection timescale of only $\sim 3$ years, this process is supposed to deplete the atmosphere rapidly. This justifies our assumption of a hydrostatic state, where $\Yatm=Y(x_{\rm He}^{(\rm A)})$.  A similar method was used in \citet{FH03}. For the most detailed discussion on He sedimentation we refer the reader to \citet{SS77b}.

\subsubsection{Key observational constraint}

\begin{figure}
\includegraphics[width=0.45\textwidth]{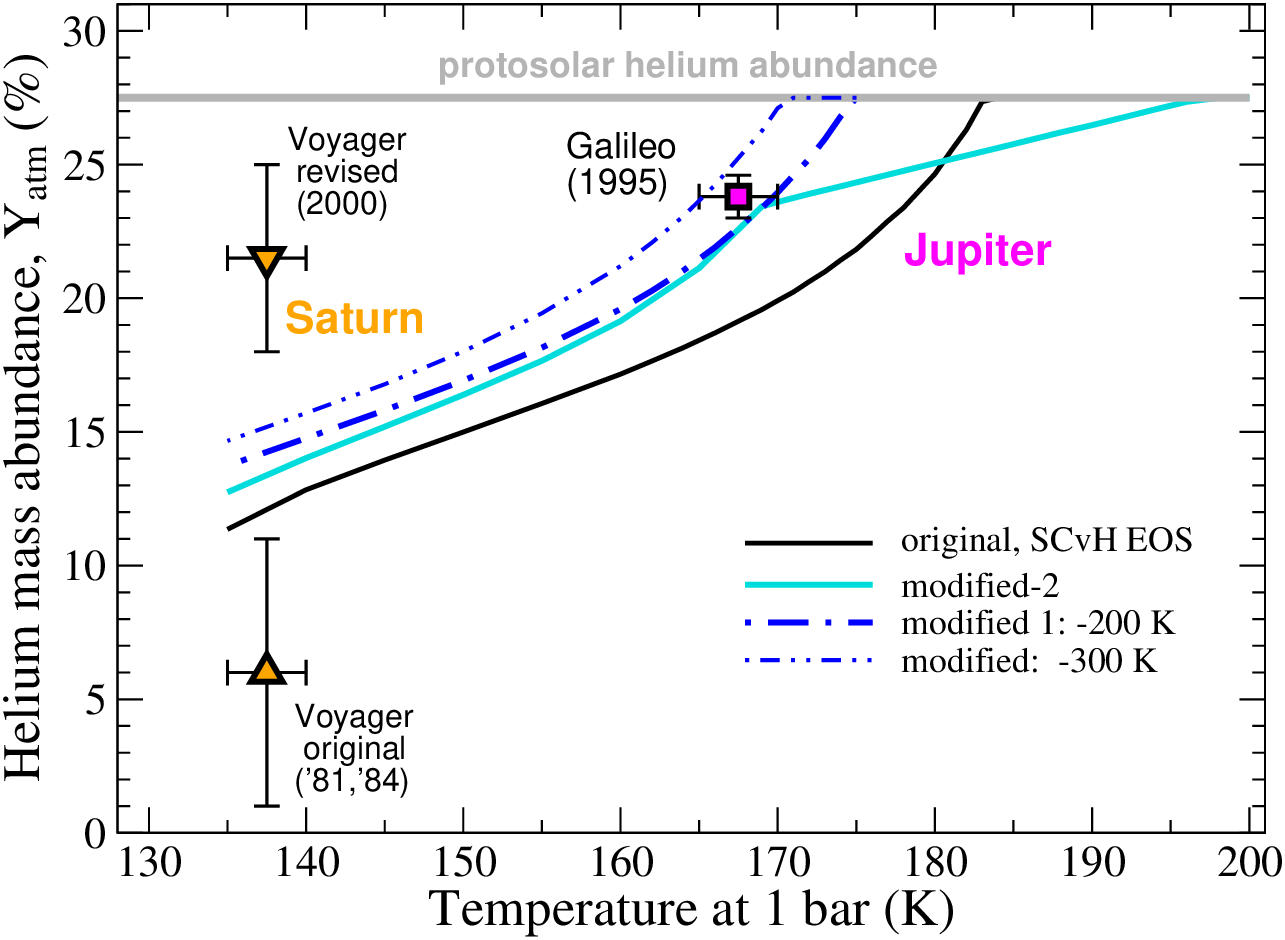}
\caption{\label{fig:Yatm_T1}
Atmospheric helium abundance $\Yatm$ in mass percent vs.~the 1-bar temperature for different constant 
offsets of the demixing temperature of the Lorenzen H/He phase diagram (\emph{blue}). An offset of -200 K (SCvH EOS),
suffices to match Jupiter's observed atmospheric helium abundance.}
\end{figure}

Because the computed value of $\Yatm$ depends on {\Tsurf}, and because {\Tsurf} decreases in the course of the planet's long-term cooling over Gyrs, the value of $\Yatm$ changes with time. In Figure \ref{fig:Yatm_T1} we show the dependence of $\Yatm$ on {\Tsurf} around the present state of Jupiter (165--170 K) and Saturn (135--140 K). Clearly, $\Yatm$ decreases with \Tsurf: colder H/He adiabats have a wider crossover with the demixing curve
at given $x_{\rm He}$, and thus require lower converged $x_{\rm He}^{(\rm A)}$ values to reach the equilibrium-point. 
The Lorenzen data yield $\Yatm=0.185$--0.20 for Jupiter (\emph{solid black curve}). This is lower than the 
\emph{Galileo} probe observational value of $Y_{\rm Jup}^{\rm (obs)}=0.238 \pm 0.005$ \citep{Zahn98}.  As we consider this measurement to be a key observational constraint on He rain in Jupiter and the H/He phase diagram, we modify the H/He phase diagram to match the data point.

\subsubsection{Modifications to the Lorenzen H/He data}

\paragraph*{Modified-1 H/He data}
As mentioned above there is considerable uncertainty about the correct demixing diagram, with differences of up to 1000 K obtained by different groups. Therefore, we introduce two modifications of the Lorenzen H/He data. In our \emph{modified-1}version,  we apply a constant temperature shift $\Delta T_{\rm dmx}$ to shift the whole H/He phase diagram according to
\begin{equation}\label{eq:deltaTdmx}
	T_{\rm dmx}^{(mod\mbox{-}1)}(P, x_{\rm He}) = T_{\rm dmx}^{\rm (Lor)}(P, x_{\rm He}) + \Delta T_{\rm dmx}
\end{equation}
until the computed $\Yatm$ value for present Jupiter is within the observational error bars of both $Y_{\rm atm}^{(\rm obs)}$ and \Tsurf. As can be read from Figure \ref{fig:Yatm_T1}, a good match is achieved by $\Delta T_{\rm dmx}=-200$ to $-300$ K, if using the SCvH-EOS, implying that perhaps the Lorenzen demixing diagram slightly under-estimates the real demixing temperatures. For our modified-1 version to the \citet{Lorenzen09,Lorenzen11} phase diagram data we use $\Delta T_{\rm dmx}=-200$ K. The phase diagram then predicts $\Yatm=0.2338$ for \Tsurf $=169$ K. 

\begin{figure}
\includegraphics[width=0.40\textwidth]{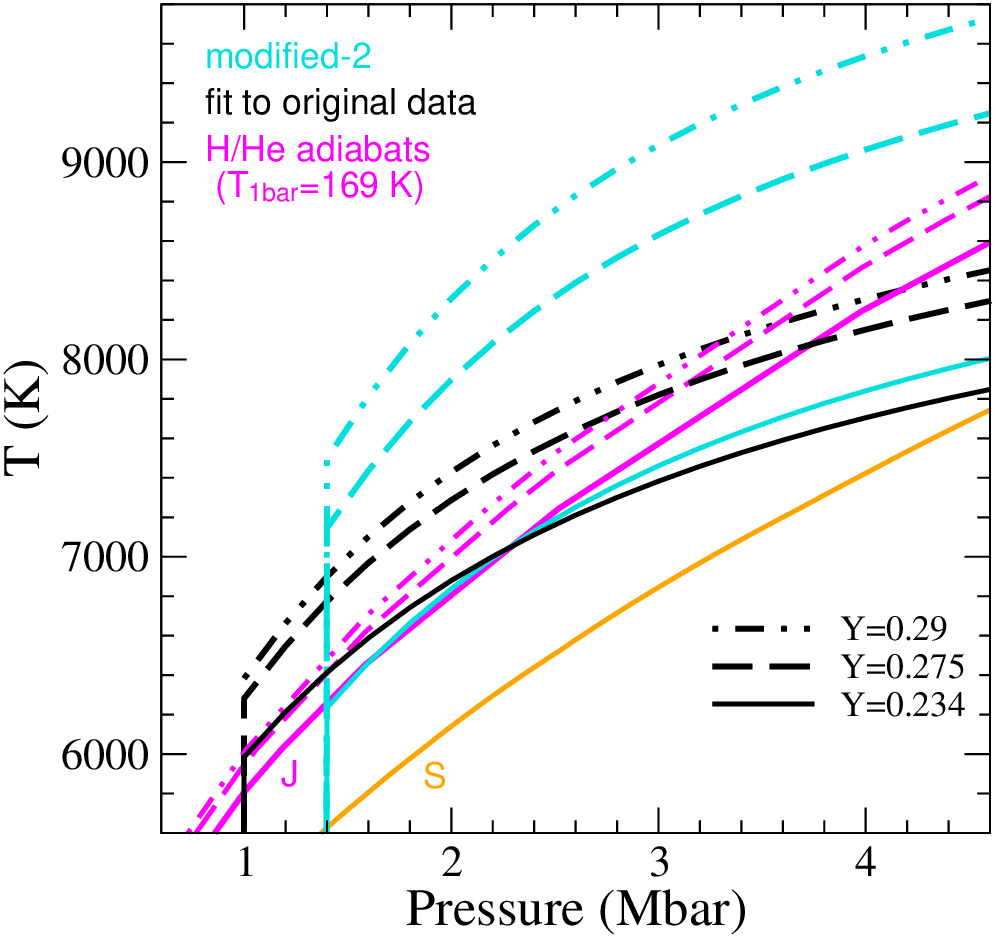}
\caption{\label{fig:dmx_mods}
Our modified-2 version (\emph{cyan}) to the (fitted) Lorenzen-H/He data (\emph{black})
for three He abundances relevant for Jupiter's interior, as can be seen by comparing to adiabats (\emph{magenta}). 
To good approximation, the \emph{solid magenta} adiabat is a Jupiter adiabat, while the \emph{orange curve} is a Saturn adiabat.}
\end{figure}

\paragraph*{Modified-2 H/He data}
In our \emph{modified-2} version of the \citet{Lorenzen09,Lorenzen11} data for H/He demixing, which was found to be driven by metallization of hydrogen, we apply a modest pressure-shift of those data by  0.4 Mbar, in which case demixing would not occur below 1.4 Mbar.  Our ad-hoc shift is inspired by the recent revision of the predicted coexistence line of the plasma-phase-transition of hydrogen toward 1 Mbar higher pressures, with now excellent agreement between the ab initio simulations of \citet{Morales13b} and the earlier shock compression experiments by \citet{Weir96}. 
In addition, we stretch $T_{\rm dmx}(x_{\rm He}, P)$ for $x_{\rm He} > x_{\rm He}(\Yatm)$ so that demixing already starts right below \Tsurf $=200$ K and proceeds with a shallower gradients in both $\Yatm$ (\Tsurf) and in $x_{\rm He}(P)$. Mathematically, we apply the modification
\begin{multline}\label{eq:mod2}
	T_{\rm dmx}^{(mod\mbox{-}2)}(P, x_{\rm He}) =  \left(T_{\rm dmx}^{\rm (Lor)}(P - 0.4/\mbox{Mbar}, x_{\rm He})\right.\\
	+ \left.\Delta T_{\rm dmx}\right)  \times \left[1 + a(x_{\rm He}-x_{\rm He,atm} )\right]
\end{multline}
with $a=0.6$ if $x_{\rm He} > x_{\rm He,atm}$, otherwise $a=0$. We chose $\Delta T_{\rm dmx}=+250 K$ to obtain $\Yatm=0.2344$ for \Tsurf $=169$ K. The modified-2 version is displayed in Figure \ref{fig:dmx_mods} for relevant He abundances, along with H/He adiabats for the surface temperature of present Jupiter.

\subsection{Internal Helium abundance profile $Y(P,T)$ }\label{sec:Yprofil}

Once the values of $\Yatm$ and the onset pressure are known from the procedure described in 
Section \ref{sec:Yatm}, we can derive the internal helium profile.

We first compute a three-layer model with outer envelope He abundance $Y_1=\Yatm$ and inner envelope He abundance $Y_2$, the latter one adjusted to conserve the total mass of helium, where the layer boundary pressure is set equal to the onset pressure of demixing, $P_{0}=1$ Mbar. This is our reference model 3L-qha-T169. We then use the inner envelope adiabat of constant He abundance $Y_2$ as the background state $T(P)$ upon which the first inhomogeneous He profile according to the demixing diagram is computed. 
The local equilibrium abundances $x^{\rm(A)}_{\rm He}$ are determined in dependence on the local temperatures and pressures along that adiabat by solving Eq.~(\ref{eq:fitTdmx}) for $x_{\rm He}(T_{\rm dmx}=T(P), P)$ using Eq.~(\ref{eq:deltaTdmx}). At that point we make use of our analytic fit to the Lorenzen data in order to obtain a smooth He gradient. 

The inner edge of the demixing region is found by requiring He mass conservation. Starting at 1 Mbar 
and working inward, we ask at each pressure level whether the integrated helium mass above, plus the proposed 
He mass under a constant extension of the local He abundance down to the core, would match the given total 
He mass. If so, the $Y$-profile is forced to leave the equilibrium curve at that mass level, $m_{23}$, and to continue with that abundance down to the core. The final internal He profile thus requires three layers: a He-poor outer envelope of $Y_1=\Yatm$ between 1 and $10^6$ bar, a demixing region with inhomogeneous He abundance between $m_{12}=m(1$ Mbar) and $m_{23}$, and an inner envelope with $Y_3>Y_{\rm proto}$ between $m_{23}$ and the core. This procedure only works as long as an $Y_3$ value can be found. Otherwise, He layer formation on-top of the core would naturally occur; see also \citet{FH03}.

The He abundances in the inhomogeneous region, the extent of the demixing region, [$m_{12}$--$m_{23}$] and the $Y_3$ value all depend on the $T(P)$ profile in the demixing region. We account for that dependence by an \emph{outer loop} which iterates between $Y(P)$, $m_{23}$, $Y_3$ on the one hand and the $T(P)$ profile on the other hand, see Section \ref{sec:outerloop}.

\subsection{Calculating the $R$-parameter}\label{sec:Rparam}

We compute the composition and temperature gradients, defined as
\begin{equation}
	\nabla_{\mu}:=\frac{P}{\mu}\frac{d\mu}{dP}\quad, \quad\nabla_{T}:=\frac{P}{T}\frac{dT}{dP}\:,
\end{equation}
where none of the thermodynamic variables are kept constant. These gradients are \emph{average} gradients 
in a sense that the average is assumed to be taken across several layers (if LDD convection occurs), 
and \emph{local} in a sense that the temperature gradient and the composition gradient may change 
over large distances of $\sim 0.1 R_{\oplus}$ (700 km). However, we never explicitly compute an average 
over layers (as LC12 do) because we to not explicitly distinguish between diffusive interfaces and 
convective, adiabatic layers (as LC12 do). 

To compute $\nabla_T$ we decompose it into the adiabatic gradient, $\nabla_{ad}(P,T,x_{\rm He})$ plus an analytically added free value. In fact, we choose the local superadiabaticity $\nabla_T-\nabla_{ad}$ as the running free parameter.  To obtain $\nabla_{ad}(P,T,x_{\rm He})$ and the local derivatives $\alpha_T$ and $\alpha_{\mu}$ we create local EOS tables around $(P,T)$ of local composition $Y(P)$. To compute $\alpha_T$ we use Equation (43) in \citet{SCvH95} and for $\nabla_{ad}$ we use Eqs. (45--46) therein but with the corrections $S/S_H\to S_H/S$ and $S/S_{\rm He}\to S_{\rm He}/S$.

For computing $\nabla_{\mu}$, we assume a given (superadiabatic) temperature profile $T(P)$ and a given mean molecular weight profile $\mu(P)$. The latter one is calculated based on $Y(P)$ as described in Section \ref{sec:Yprofil}, and by using $\mu^{-1}=X \mu_{\rm H}^{-1} + Y\mu_{\rm He}^{-1}$, where $X$ and $Y$ are the mass fractions of H and He, respectively. The computed composition gradient $\nabla_{\mu}$ describes the average gradient under our assumption of instantaneous He sedimentation. We also apply it to the computation of the density ratio $\Rinv$, an approximation that leaves room for future explorations of the physical processes involved with He rain, and deserves some discussion.  

The density ratio $R_0$ is thought to express the buoyancy experienced by a vertically displaced parcel in a sourrounding that may have temperature ($\nabla_T$) and composition ($\nabla_{\mu}$) background gradients, like the ones we computed here. 
In mixing length theory for convection, it is generally assumed that a vertically  displaced parcel expands/contracts adiabatically and maintains its composition because diffusion of particles and of heat occur on longer time-scales than convective transport. Here we face a different situation. In the ODD regime, diffusion of heat and particles out of the parcel may occur, see \citet{SS77b} about \emph{overstable modes}. 
Morover, our assumption of \emph{instantaneous} He sedimentation implies rapid He condensation, 
so that droplets, if formed in the parcel, may leave it and thus alter its composition during the journey, contrary to our fundamental assumption ($iii$). This effect would tend to reduce the composition difference between the moving parcel and its ambient fluid. We can account for this possible reduction by introducing a factor $\beta\:\epsilon\: [0,1]$, so that the relevant density ratio that determines the stability of the system becomes
\begin{equation}
	\Rinv = \frac{\alpha_{\mu}}{\alpha_T}\:\frac{\beta\nabla_{\mu}}{\nabla_T-\nabla_{ad}}
\end{equation}
and $\nabla_{\mu}$ is the background composition gradient as described above. 

The end-member case $\beta=1$ (our fundamental assumption $iii$) reflects the usual assumption of conserved composition. 
A value $\beta \lesssim 1$ may apply if the time-scale for the formation of initial He droplets is longer than the eddy life-time, so that He droplet formation in a vertically moving parcel becomes a rare event at first place. Still, some droplets may form and sediment out, however, so that $\beta < 1$. 
In fact, $\beta=1$ might be inconsistent with our assumption of instantaneous semimentation. If applied to a parcel,
the He abundance therein would equal that dictated by the phase diagram for the parcels' own temperature and thus tend to decrease when it moves upward. \citet{SS77b} suggest $1-\beta < 0.97$. In fact, $\beta \ll 1$ will be preferred according to our results. We note that the real composition gradient in the planet is not well known. Heavy elements may contribute to a stabilizing gradient. For instance, the Jupiter models by \citet{N+12} predict an increase in heavy element abundance at $P\geq 4$ Mbar, and the most recent ones even at $P\geq 3$ Mbar \citep{Becker+14}, which is located within or near the 
lower end of the demixing region. For simplicity we apply $\beta=1$ in this work, keeping in mind that we may 
over-estimate the stabilizing He gradient between the rising parcel and its surrounding. 
A physically self-consistent treatment of the composition gradient is left to future work, see also Section \ref{sec:Earth}.

\subsection{Consistency between Y(T) and T(Y): the outer Loop}\label{sec:outerloop}

\begin{figure}
\includegraphics[width=0.40\textwidth]{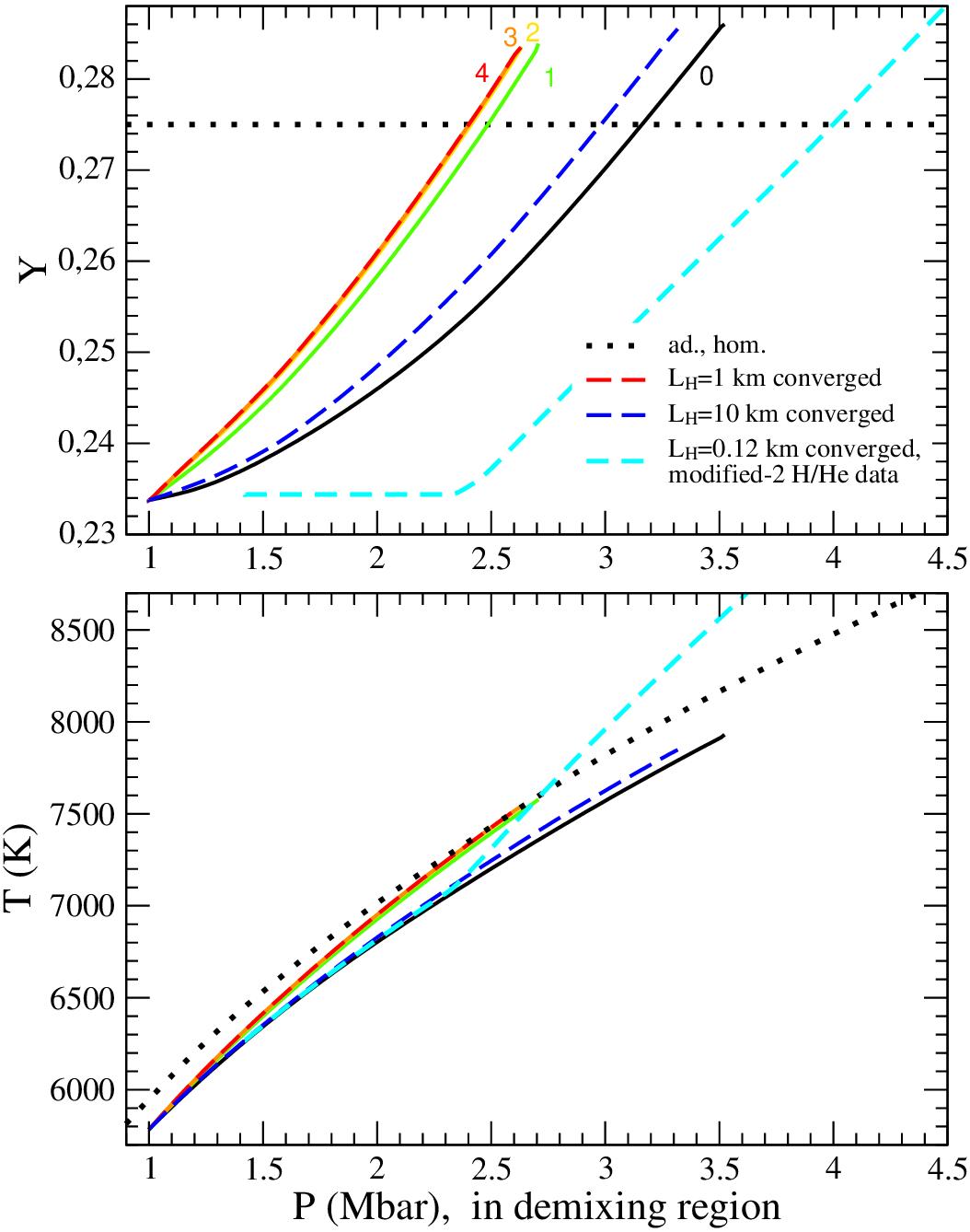}
\caption{\label{fig:iterYT}
Convergence of He abundance profile (\emph{upper panel}) and temperature profile (\emph{lower panel}) toward a solution that is consistent with respect to $Y(T,P)$ and $T_{dmx}(P,Y)$ shown for the case of $L_H=1$ km and the modified-1 H/He data. The outer loop begins with an quasi-homogeneous adiabat (\emph{black solid}). The locally increasing He abundances leads to locally larger temperatures to transport the heat flux. 
For $L_H=1$ km (\emph{red curve}), the superadiabaticity of the converged solution leads to the same internal temperatures at the end of he demixing region as in case of no demixing (\emph{black dotted}), while for $L_H=10$ km (blue dashed) to smaller ones, and for $L_H=0.12$ km in combination with the modified-2 H/He phase diagram (\emph{cyan}) to higher ones.}
\end{figure}

The temperature profile that is needed to transport the heat flux in the presence of a composition
gradient depends on that gradient, in our case the helium abundance profile, while the
latter one at the H/He demixing--mixing phase boundary depends on the local temperatures.
To ensure consistency between $T$ and $Y$ we iterate between the $T$-profile for a given $Y(P)$-profile and the 
$Y$-profile for given $T(P)$. After at most 5 iterations good convergence
is achieved. This is illustrated in Figure \ref{fig:iterYT} for the case of $l_H$=1 km. The iteration starts with the 3L-qha-T169 model adiabat (\emph{solid black} curve). This adiabat is colder than the fully homogeneous adiabat of $Y=Y_{\rm proto}$ (\emph{dotted} curve) because He-poor regions as at the outer boundary require lower temperatures for
maintaining constant entropy. Superadiabatic temperatures (\emph{coloured} curves) require higher He abundances for consistency with the H/He demixing curve, see Figure \ref{fig:Tdmx}.  In turn, higher He abundances for a fixed 
\Yatm\ value imply a steeper He gradient, and thus need higher temperatures for transporting the heat flux. 
Convergence is rapid for the moderate superadiabaticities in LDD convection.

As a test case we also imposed the constraint of overturning convection ($\Rinv=1$), in which case the converged He abundance by the end of the demixing region was found to be higher than allowed by He mass conservation.  In other words, the condition $\Rinv=1$ can only be satisfied in Jupiter by letting $\nabla_{\mu}$ and $\nabla_{T}$ go to infinity, meaning a step in $Y$ and $T$. We here recover the runaway effect between the gradients in $Y$ and $T$ as observed by \citet{FH03}.

Having put the pieces together, we can compute the effect of LDD and ODD convection 
in the demixing region on Jupiter's present structure.

\section{Results for the Modified-1 H/He demixing diagram}\label{sec:res1}

In this Section we apply the modified-1 H/He demixing phase diagram to models of Jupiter and assume that 
either ODD or LDD convection occurs in the demixing region with the layer height as a free parameter.

\subsection{ODD convection}\label{sec:res1ODD}

\begin{figure}
\includegraphics[width=0.45\textwidth]{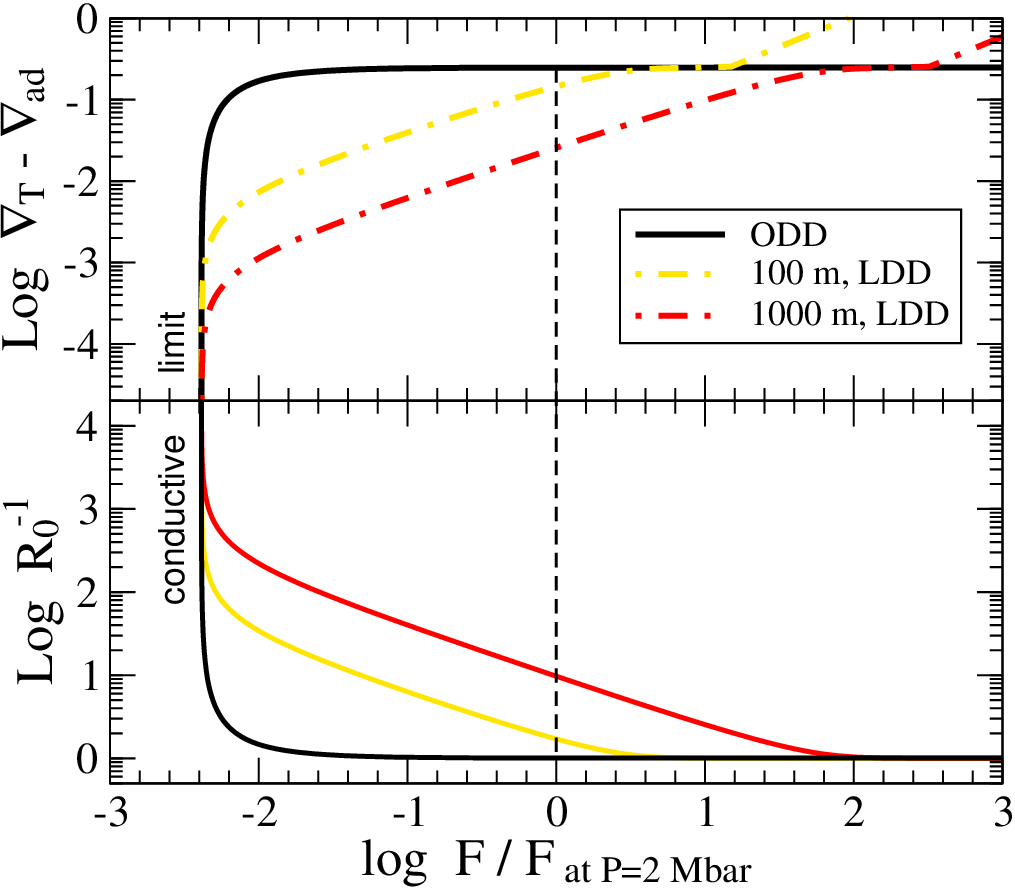}
\caption{\label{fig:ODD}
Relation between superadiabaticity, density-ratio, and total heat flux for parameter values in Jupiter's demixing region at 2 Mbar. The values of $R_0^{-1}$ and $\nabla_T-\nabla_{ad}$ that will be assigned to the 2 Mbar pressure level for a given layer height of respectively 100 m (yellow) or 1000 m (red), or in the case of ODD convection (solid black) are those that cross the black dashed line. For ODD convection, a sufficiently high heat flux can only be obtained for $\Rinv\simeq 1$, which violates the assumption of ODD.
The relations are displayed here for the initial iteration step, i.e. for a He gradient along the 3L-qha-T169 adiabat.
The curves do not display final models for Jupiter, as discussed in the text.
}
\end{figure}

ODD and LDD convection can lead to very different resulting superadiabaticities and density-ratio values in Jupiter's demixing region at a given pressure level, for instance at 2 Mbar as shown in Figure \ref{fig:ODD}. When the running free parameter $\nabla_T-\nabla_{ad}$ is low ($<0.1$), $\Rinv>1$. In the case of ODD convection the factor $(\Rinv-1)^{-1}$ in Eqs.~(\ref{eq:NuT_ODD}) then yields a flux too low, close to the diffusive limit. In fact, in ODD convection the flux reaches the order of $F_{\rm Jup}(P)$ only for $(\Rinv-1)^{-1}\to\infty$, i.e.\ for $\Rinv\simeq 1$, and this behaviour occurs over the entire demixing region. A value $\Rinv\simeq 1$ indicates the preference for overturning instead of semi-convection. We conclude that in order to transport the given given heat flux by ODD convection, the required superadiabaticity would be so high that the system would want to transition to overturning convection, which contradicts the assumption of ODD convection.  Therefore, this does not appear to be a viable path towards a Jupiter model.

At a given superadiabaticity, the heat flux that can be transported by LDD convection is 1--2 orders of magnitude higher than in ODD convection, even if the layer height is quite small ($<1$ km). LDD convection thus requires lower superadiabaticities. Thus we focus on LDD convection.

\subsection{LDD convection}\label{sec:res1LDD}

\begin{figure*}
\includegraphics[width=0.7\textwidth]{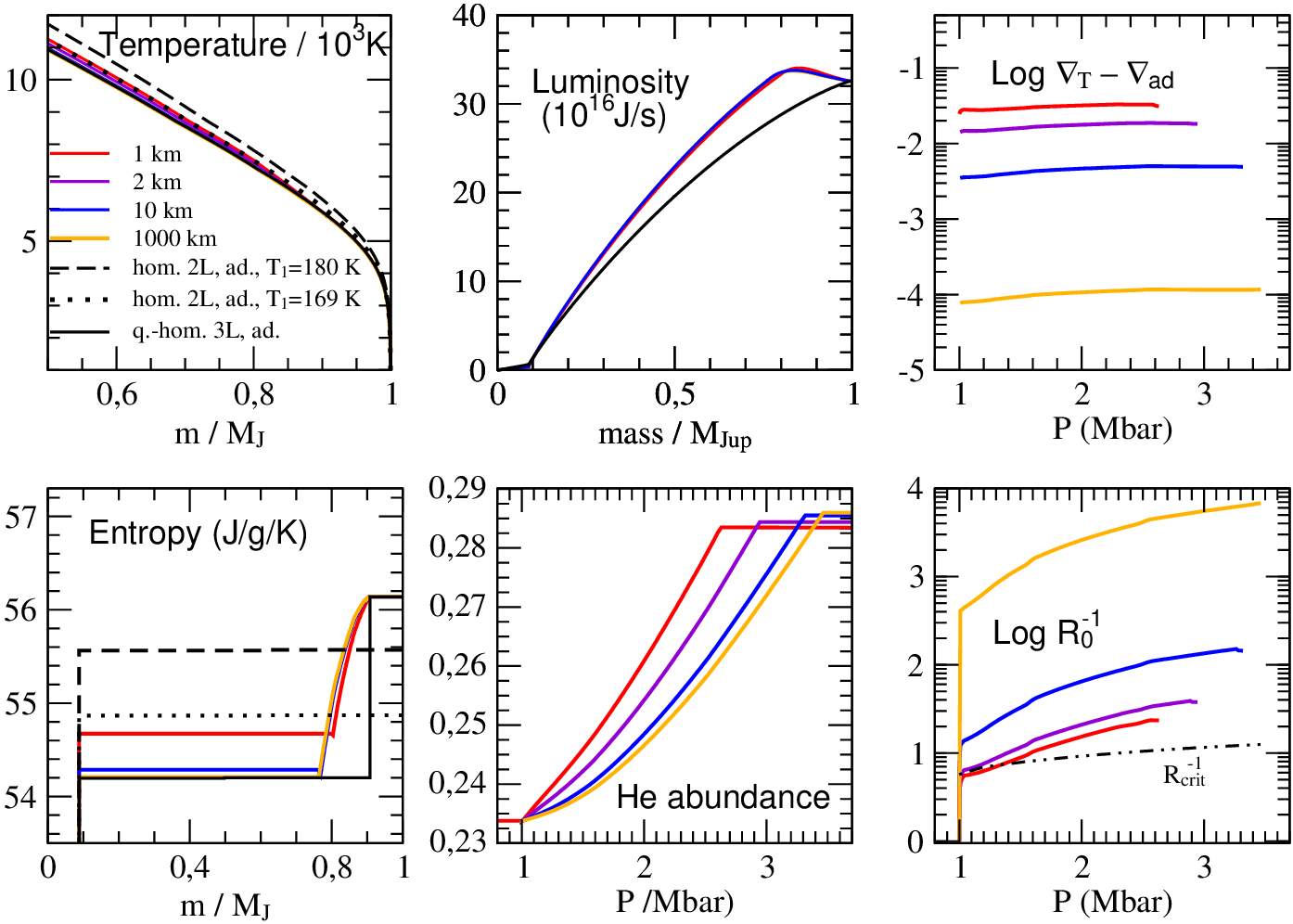}
\caption{\label{fig:dmx6} Fully converged resulting internal profiles in and around the H/He demixing region of models for present Jupiter for  different assumed heights of LDD convection\emph{(colour coded)}, and using the modified-1 H/He data. See running text and Legend in upper left panel for description. 
Note that these models are shown for illustration ; all but the \emph{red} one can be ruled  out.}
\end{figure*}

Figure \ref{fig:dmx6} shows the resulting profiles of temperature, specific entropy, luminosity, heat flux, helium abundance, superadiabaticity, and $R$-parameter in Jupiter's demixing region for various assumed layer heights between 1~km and 1000~km (coloured curves). Figure \ref{fig:dmx6} also shows three black curves. The \emph{black dashed} curve is for the 2L-ha-T180 model (see Section \ref{sec:refmodels}). The \emph{black dotted} curve is for the 2L-ha-T169 model, and is supposed to describe the present Jupiter if demixing would never have occurred. Finally, the \emph{black solid} curve is the 3L-qha-T169 model and is used as the initial state in our double-iterative procedure. All these models have a $28\:\ME$ rock core and a pure H/He envelope.

\paragraph*{Entropy}
In the outer part of the demixing region and in the adiabatic outer envelope, the entropy is seen to rise above the level of the \Tsurf=180 K reference state before demixing began. Therefore, that outer part gives a negative contribution to the total luminosity. The rise above the reference state might surprise, in particular as \citet{FH03} find (for Saturn) the entropy in the outer part to decrease steadily with \Tsurf, see their Figure~7. We argue that this difference is due to the different H/He phase diagrams used, especially due to the small {\Tsurf}-interval over which demixing occurs in Jupiter ($s$ increases with \Tsurf), and the strong He depletion ($s$ decreases with $\Yatm$). Here, the He depletion wins over the cooling effect in the time-evolution of the outer envelope's entropy.

\paragraph*{Extent}
In terms of pressures, the demixing region extends from 1 Mbar to at most 3.5 Mbar. While by definition the entropy is constant outside the demixing region, it changes steadily within it, mostly due to the steadily changing He abundance. From the entropy panel in Figure \ref{fig:dmx6} we can derive an extension of the demixing region over $dM=0.1$--0.15 \Mjup ($dM=30$--$47 \ME$, or $dR=0.92$--$1.27 \RE\sim 5,900$--8100 km), depending somewhat on the layer height, with smaller layer heights yielding thinner demixing regions.

\paragraph*{Density-ratio}
For layer heights above 1 km, the values of $\Rinv$ are all higher than $\Rcrit$. We remind ourselves that $\Rcrit$ is an upper limit that marks the transition to the diffusive regime. Values $\Rinv>\Rcrit$ are obtained as a result of low superadiabaticities (see the lower left panel in Fig.~\ref{fig:dmx6}). Under the assumption of $\beta=1$ this implies that LDD convection can transport the heat flux too efficiently if the He abundance gradient obeys the modified-1 H/He data. The larger the layer height, the fewer interfaces are present, and thus the smaller becomes the required superadiabaticity, leading to higher $\Rinv$ values. The largest $\Rinv$ values are then obtained for the largest assumed layer height, which is half the size of the demixing region ($l_H \approx 3600$ km). 
Clearly, in order to get resulting $\Rinv$ values in agreement with the regime of LDD convection as seen in numerical experiments \citep{Mirouh12,Wood13}, smaller layer heights ($l_H<1$ km) are required for present Jupiter. This finding is in agreement with what we derived from our toy models in Section \ref{sec:fT}. How small can the layer height be?

\paragraph*{Minimum Layer Height}
The shorter the layer height, the warmer the adiabatic deep interior and the higher its specific entropy; thus, the smaller becomes the entropy difference $ds(m)$ with respect to the internal profile at the previous time step. A minimum layer height is obtained when the summation over $T\,ds$ in Jupiter's deep interior is no longer capable of compensating the negative luminosity contribution from the planet's outer part, so that the total luminosity would become negative.  We find this minimum to be 1 km if the layer height is kept at constant value. This value of 1 km is imposed by the condition of a positive total planet luminosity, and \emph{neither} by the Ledoux instability criterion, which would be violated ($\Rinv<$ 1) at $l_H \approx 10$ m, nor by the minimum length scale criterion (Equation \ref{eq:dmin}).

\paragraph*{Thermal evolution}
The energy that can escape from the planet is determined by the atmosphere model. If more (less) energy is released from the interior, it will take longer (less long) to transport it through the atmosphere.
Because of additional gravitational energy from sinking He droplets, the effect of He rain is generally thought to prolong the cooling time of a planet \citep{SS77b,Saumon+92,Hubbard99,FH03}. 

\begin{figure}
\includegraphics[width=0.45\textwidth]{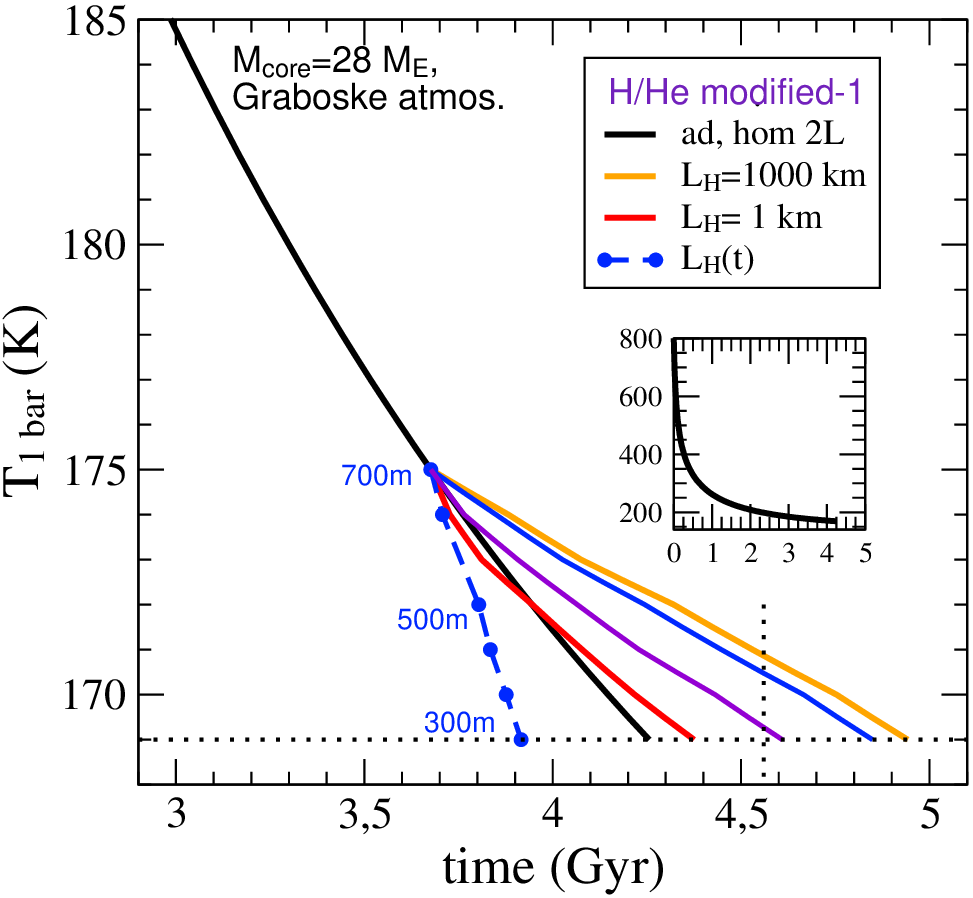}
\caption{\label{fig:evol1}Cooling curves for different layer heights as in Figure \ref{fig:dmx6} or as labelled (\emph{blue-dotted}) in comparison with adiabatic, homogeneous evolution (\emph{black}). The \emph{vertical dotted line} indicates the age of the solar system.}
\end{figure}

Figure \ref{fig:evol1} shows the effect of He rain and LDD convection on Jupiter's cooling time relative to that of homogeneous evolution for different assumed layer heights. All displayed cooling curves have a $T_{\rm eff}$ of 124.4 K within the observational error bars, and make use of the Graboske model atmosphere.
For large layer heights  (1000 km) redistribution of He dominates over the temperature effect, so that we observe the expected prolongation of the cooling time. As for $l_H=1000$ km the superadiabaticity is negligibly small, this case can be considered equivalent to the usual assumption of adiabatic cooling ($\beta=0$), where the cooling behaviour is only influenced by the additional gravitational energy from the He rain \citep{Hubbard99,FH03}. For the modified-1 H/He diagram, such a model would yield a cooling time prolongation by $\sim 0.7$ Gyrs, compare the \emph{orange} and the \emph{black} curves in Figure \ref{fig:evol1}. 

Conversely, if the deep interior of a planet is prevented from efficient cooling, less energy can escape and the cooling time will tend to decrease. This case has been suggested to apply to Uranus and to explain its faintness  \citep{Hubbard+95}. Indeed, we see that the shorter the layer height (e.g.~1 km vs.~1000 km), the shorter becomes the cooling time, as expected For $l_H=1$ km, the effects of additional gravitational energy and of inhibited heat transport balance each other, and the resulting cooling time is about the same as in the fully adiabatic, homogeneous case (compare the \emph{red} and the \emph{black} curves in Figure \ref{fig:evol1}). 

\begin{figure}
\includegraphics[width=0.47\textwidth]{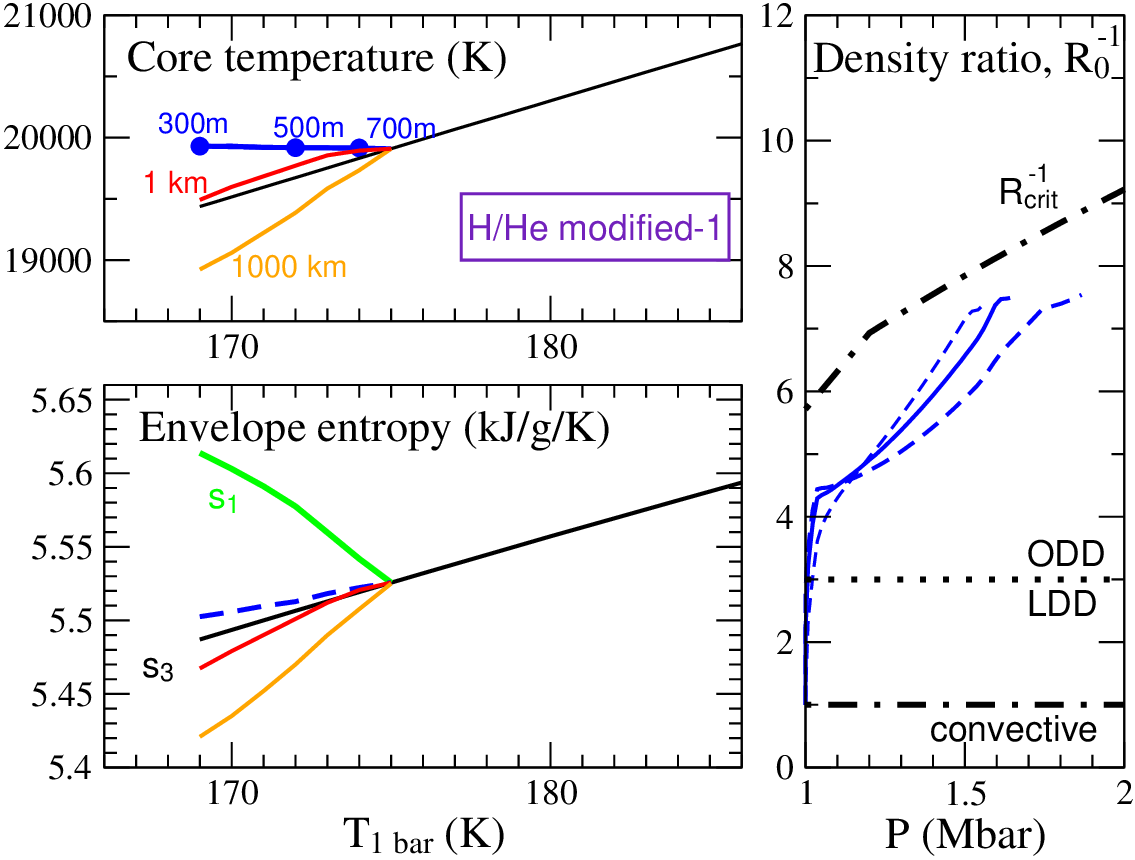}
\caption{\label{fig:TcRss1}Results for the evolution in terms of surface temperature using the 
modified-1 H/He demixing data. 
\emph{Upper panel:} core temperature; \emph{lower panel:} entropy in the homogeneous, adiabatic outer envelope ($s_1$; green) and in the inner envelope ($s_3$). Curves for different $l_H$ values are colour coded, while \emph{solid black} curves are for homogeneous, adiabatic 2-layer models; \emph{right panel:} density ratio for the models with variable layer heights over time.}
\end{figure}

For even shorter layer heights of a few 100m that are achieved by letting the layer height vary over time, cooling of the interior becomes significantly stalled so that the core temperatures may stay constant or even slightly increase over time, see Figure \ref{fig:TcRss1}. In that case, the cooling time shortens to be less than Jupiter's known age of 4.56 Gyr (see the \emph{blue-dashed curve} in  Figure \ref{fig:evol1}). 
Figures \ref{fig:evol1} and \ref{fig:TcRss1} show that a resulting cooling time of 4.6 Gyr for Jupiter will require a balance between the the decrement in deep internal entropy due to increasing He abundance and its increment due to the superadiabaticity, so that the entropy values of the adiabatic, homogeneous case are about recovered. 
We find that using the modified-1 H/He data, the balance occurs for $l_H\approx 1$ km. However, the resulting $\Rinv$  values assuming $\beta=1$ lie above the allowed range. This is an important results that we consider to be a clue to $\beta<1$. For $\beta=0.05$--0.5 ($\approx 1/\Rcrit$), the \emph{red} modified-1 H/He data based model would satisfy all constraints. This implies a reduced composition difference between a vertically moving parcel and its superadiabatic surrounding compared to the difference $\nabla_{\mu}$ we compute under our assumption ($iii$).

\paragraph*{Conclusion}
Given our fundamental assumptions, all models of this section are ruled out because they violate the $1<\Rinv<\Rcrit$ criterion. If we drop assumption ($iii$), the modified-1 H/He diagram based model with $l_H=1$ km can satisfy all constraints  and predicts $\beta=0.05$--0.5.

\section{Results for the Modified-2  H/He demixing diagram}\label{sec:res2}

In the previous Section we have seen that none of the models for the modified-1 H/He phase diagram could yield a consistent solution under our fundamental assumptions (Section \ref{sec:fundass}). For LDD convection, we obtained too large values $\Rinv>\Rcrit$, partly as a result of assuming $\beta=1$, while for ODD convection too small values $\Rinv \simeq 1$ as a result of too large superadiabaticities.  In this Section we apply the modified-2 H/He phase diagram. It has been designed to lead to an earlier begin of demixing in time and a deeper onset of demixing within the planet while as well reproducing the \emph{Galileo} probe observational value of $Y_{\rm atm}$ at the present. In addition to using the modified-2 H/He data we here assume a time-variable layer height. Our reasons why we opt for such modifications will be discussed in the following.

\subsection{ODD convection}

For the assumption of ODD convection in the demixing region we obtained too large superadiabaticities in Section \ref{sec:res1} because ODD convection was too inefficient to transport the heat flux. Of course, whether or not ODD convection is efficient enough depends on the amount of heat to be transported. With the modified-1 phase diagram, the outer envelope on-top the demixing region yielded a negative contribution to the total intrinsic luminosity because the increase in entropy there (denoted by $s_1$) due to the change in  He abundance is a larger effect than the decrease in entropy due to cooling, so that $s_1$ increased with time (i.e.~with decreasing \Tsurf), see the \emph{green curve} in the entropy-panel in Figure \ref{fig:TcRss1}. Thus the \emph{deep interior} had to provide a large internal heat flux, even slightly more than the large, observed flux.

\begin{figure}
\includegraphics[width=0.47\textwidth]{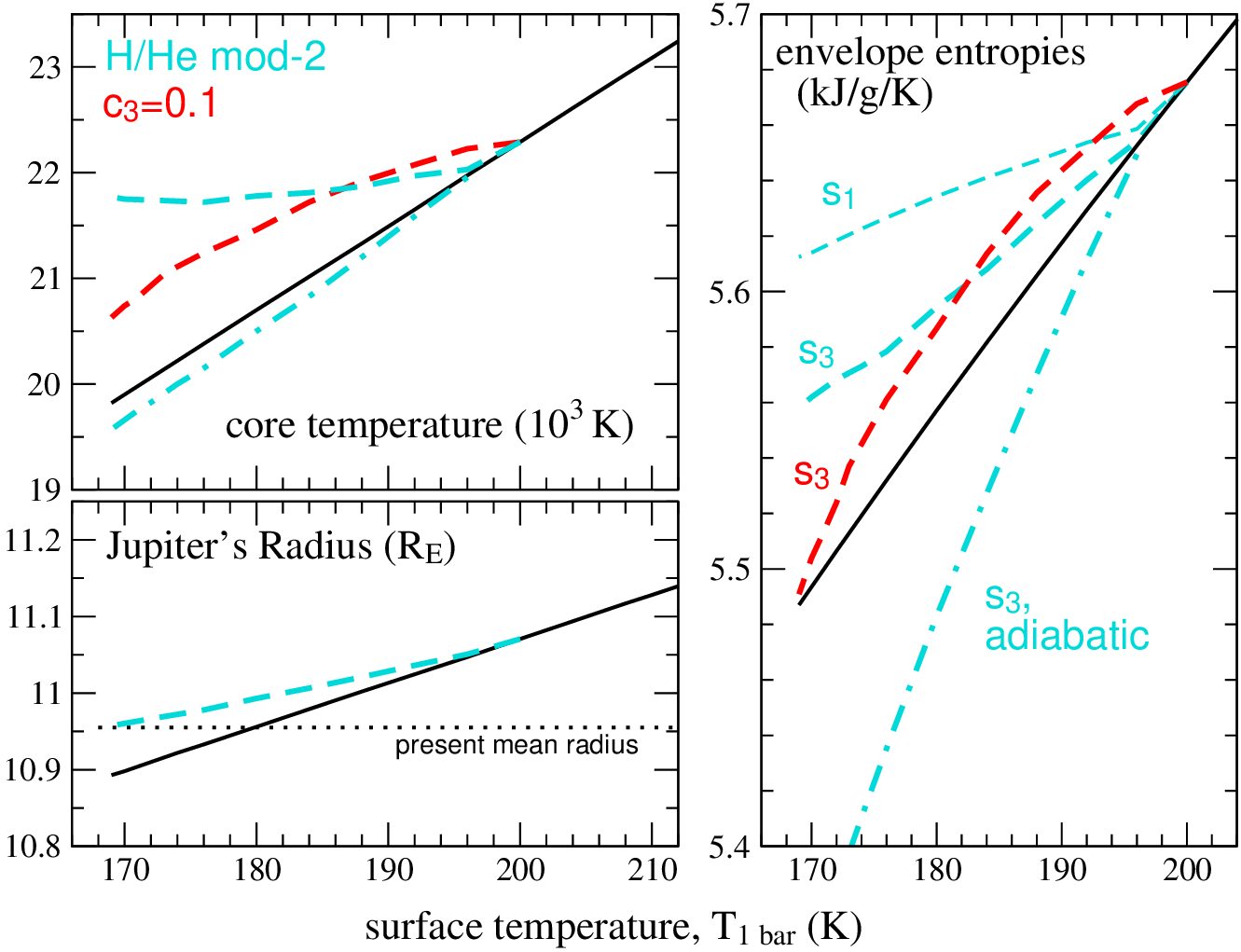}
\caption{\label{fig:TcRpss2}
Results for the evolution of core temperature, planet radius, and envelope entropies in terms of {\Tsurf}, which decreases with time, using the modified-2 H/He demixing data.
\emph{Solid black} curves are for homogeneous, adiabatic evolution; \emph{dashed cyan} curves for the modified-2 H/He phase diagram and LDD convection; \emph{red} curves for a toy composition gradient $c_3=0.1$ and LDD convection, while \emph{dot-dashed cyan} curves are for the modifed-2 H/He data but imposed zero-superadiabaticity.}
\end{figure}

However, if the observed total luminosity would instead result from the cooling of the \emph{outer} envelope while heat release from the deep interior is strongly impeded due to the presence of an ODD or stable region,  a solution with ODD convection may exist. To invert the sign of the luminosity contribution from the outer envelope, He rain would have to proceed more slowly in time so that the effect on the entropy from the cooling of the outer envelope wins over the effect of the He depletion. This is one of the reasons why we have introduced the factor $a$ in the modified-2 H/He phase diagram, so that the He rain begins already at {\Tsurf} $=200$ K, rather than at 175 K as predicted by the modified-1 phase diagram (Figure \ref{fig:Yatm_T1}). Yet, although the modified-2 H/He phase diagram leads to the desired sign-change for the luminosity contribution from the outer part, as can be concluded from the decrease of $s_1$ with decreasing {\Tsurf} in the entropy-panel of Figure \ref{fig:TcRpss2}, we were not able to find a solution with the desired shut-down of the deep internal heat flux. At the current stage, we do not know whether a solution for ODD convcetion exists at all. Anyway, a nearly stably stratified deep interior would impose a challenge to the generation of the observed magnetic field. In the following we therefore consider LDD convection.

\subsection{LDD convection}

Contrary to the above described ODD case, in the LDD case we want the superadiabaticity to become \emph{higher}, namely by a factor of a few compared to the results of Section \ref{sec:res1LDD}. As we have seen in Figure \ref{fig:dmx6}, one way to achieve this is to shrink the layer height; but we have also seen that the minimum possible layer height is limited by the condition $L_{tot}>0$.  In fact, $L_{tot}<0$ happens if the deep internal temperatures get too high at first place, in which case the deep internal entropy (labelled $s_3(T,P)$) would \emph{increase} with time ($\delta q(m)>0$ in Eq.~\ref{eq:lm}), resulting in a huge negative luminosity contribution from the mass-rich deep interior that is impossible to be compensated for by the cooling of the outer envelope.   What we therefore need in order to enable higher superadiabaticities, is a higher upper limit on the allowed $s_3$ value, or almost equivalently ($s$ is an increasing function of $T$), on the core temperature.
To a good approximation, the upper limit on the core temperature in the presence of H/He demixing is given by the core temperature before demixing begun to operate\footnote{A tiny enhancement of core temperature with time could still be allowed for because $s_3$ also is a decreasing function of He abundance, which increases with time. \citet{SS77b} even suggest a strong heating up of the planet for the case of inhibited heat transport through the He rain region.}. Thus if we let H/He demixing start earlier in the evolution, the core temperature at that time will have been higher. This is another reason why we have stretched $T_{\rm dmx}(x_{\rm He})$ by inserting the factor $a$ in the modified-2 H/He phase diagram (Eq.~\ref{eq:mod2}).  
For our illusrative calculations with the modified-2 data we assume a core mass of $32\: \ME$, which yields a 
present-day planet radius in good agreement with Jupiter's observed one  (Figure \ref{fig:TcRpss2}).

\begin{figure}
\includegraphics[width=0.45\textwidth]{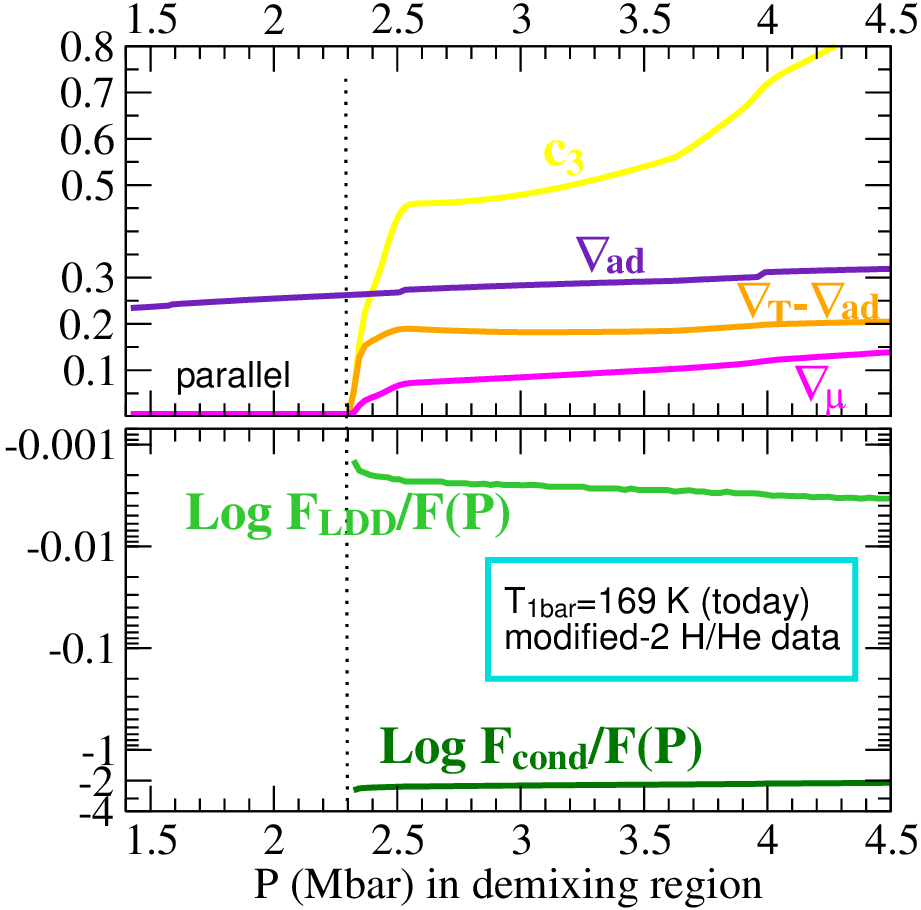}
\caption{\label{fig:fluxes} Gradients and fluxes in the demixing of Jupiter computed using the modified-2 H/He phase diagram.
Upper panel: Mean molecular weight gradient (\emph{magenta}), adiabatic gradient (\emph{violet}), 
superadiabaticity (\emph{orange}), and $c_3=\nabla_{\mu}(\alpha_{\mu}/\alpha_{T}$); 
lower panel: Heat flux contribution $F_{\rm LDD}/F$ (\emph{green}), heat flux contribution $F_{cond}/F$ (\emph{dark green}). Between 1.4 and 2.3 Mbar, the Jupiter adiabat runs \emph{parallel} to the H/He modified-2 phase diagram.}
\end{figure}

Figure \ref{fig:TcRpss2} shows the evolution of $\Tcore$ and of the envelope entropies $s_1$ and $s_3$ in terms of {\Tsurf}, which decreases with time. Indeed, the modified-2 H/He phase diagram allows for $\sim 2000$ K higher core temperatures, implying higher possible superadiabaticities in the demixing region, than the modified-1 H/He phase diagram did (Figure \ref{fig:TcRss1}). Figure \ref{fig:fluxes} shows $\nabla_T-\nabla_{ad}\approx 0.2$ for the modified-2 phase diagram case, while the highest value we could obtain for our models using the modified-1 phase diagram (and constant layer height over time) was $\approx 0.035$, see Figure \ref{fig:dmx6}. 

In order to maintain high superadiabaticities during the evolution, we adjust at each time-step (practically, at each \Tsurf-step) the layer height to be the smallest possible one within 20 per cent that does not lead to a violation of $L_{tot}>0$ or of the minimum length scale criterion. It turns out that the layer heights would decrease with time starting at about 1 km at the beginning of demixing, and ending up to be 100-200m at present. This result agrees with what we have learned from our toy models in Section \ref{sec:fT}.
 
\subsubsection{Consistency with $\Rinv$} \label{sec:Rinv2}

We next turn to the question whether our modified-2 H/He diagram based models are consistent with the range of allowed $\Rinv$ values of LDD convection. 

\citet{Mirouh12} have in detail investigated the  range of $\Rinv$(Pr,$\tau$) where layer formation occurs. In their simulations they see it to develop rapidly for $\Rinv$ values close to the overturning instability limit ($\Rinv=1$);  they also see layers to emerge for larger values of $\Rinv=$1.5--2 but only after a long simulation time, and never observe layer formation for $\Rinv > 2$. Complementary, \citet{Mirouh12} also investigate the layer formation regime according to the $\gamma$--instability theory of \citet{Radko03}. Transferred to a semi-convective system where the temperature gradient acts destabilizing and the composition gradient stabilizing, that theory predicts that an ODD system  in which $\Rinv$ is systematically lowered will eventually develop convective layers when $\gamma^{-1}$ becomes a decreasing function of $\Rinv$. The crucial parameter in this theory, $\gamma^{-1}$, depends on the composition and thermal fluxes, which can be measured during the simulations. Our point of interest here is that the $\gamma$-instability theory predicts a wider range of $\Rinv$ values for layer formation than observed in the simulations. For the Pr and $\tau$ number values relevant to Jupiter, the transition between LDD and ODD convection occurs at $\Rinv\approx 3$, with $1<\Rinv\leq 3$ being the LDD regime. If expressed in terms of the stratification parameter $r=(\Rinv-1)/(\Rcrit-1)$, the transition occurs at $r\approx 0.2$ with $0<r<0.2$ being the LDD regime. 

\begin{figure}
\includegraphics[width=0.45\textwidth]{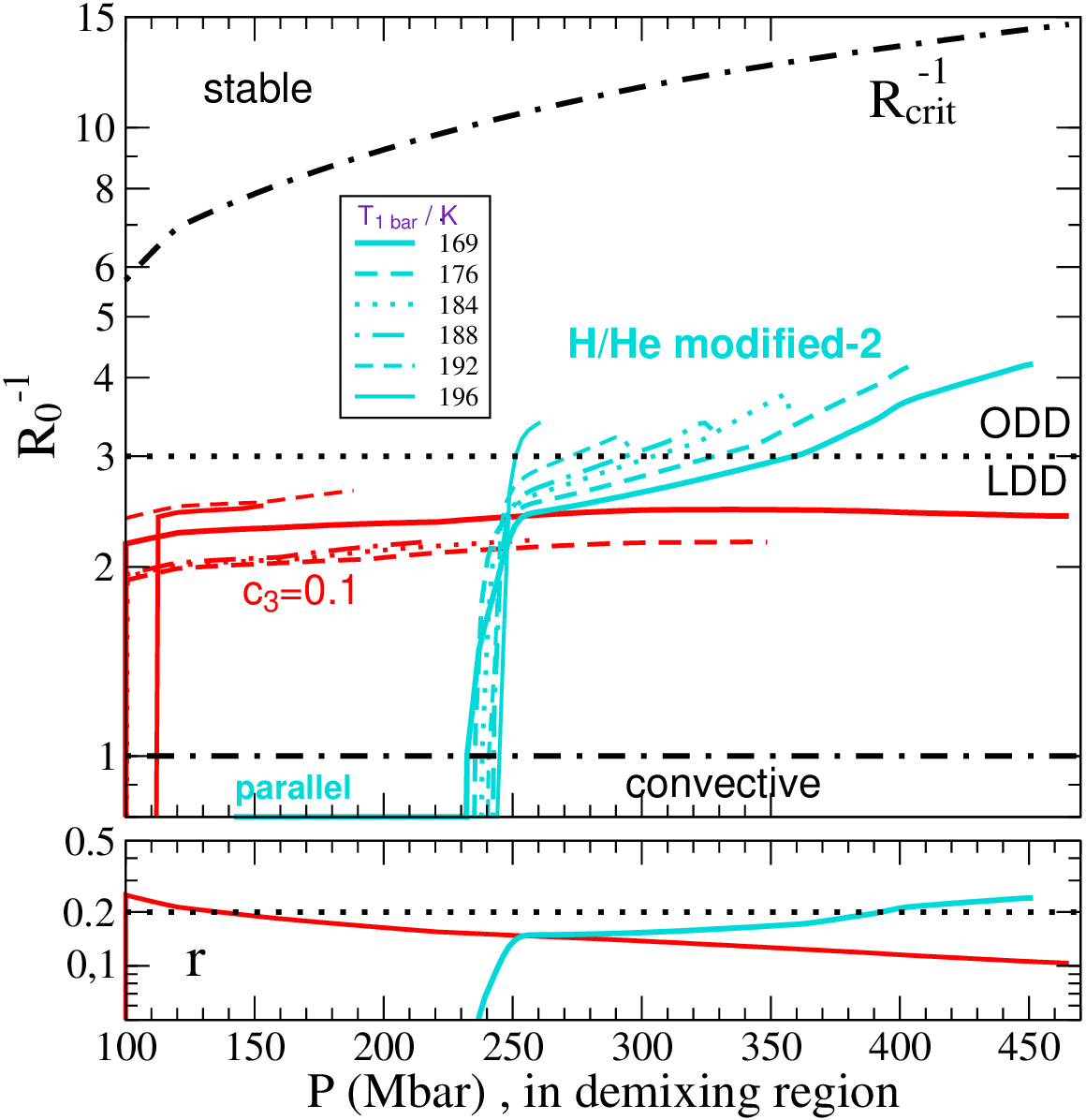}
\caption{\label{fig:Rparam}
\emph{Upper panel}: Resulting $\Rinv$ values in Jupiter's He rain zone at different {\Tsurf} values during evolution using
the modified-2 H/He phase diagram (\emph{cyan}) or $c_3=0.1$ (\emph{red}). 
\emph{Lower panel}: stratification parameter $r=(\Rinv-1)/(\Rcrit-1)$. }
\end{figure}

Figure \ref{fig:Rparam} shows the resulting $\Rinv$ values for a modified-2 H/He data based model at various {\Tsurf} values during the evolution.  
At all times, about half of the demixing region has resulting $\Rinv$ values between 1 and 3 and $r<0.2$, in agreement with the assumption of LDD convection. Although in the other half of the demixing region the $\Rinv$ values increase up to a value of 4 ($r$ up to 0.3) and thus slightly excess the upper limit of 3, further fine-tuning of the layer height through allowance of variation with depth might bring the resulting $\Rinv$ values in full agreement with the allowed range of 1--3 throughout the whole demixing region, so that we consider this \emph{cyan} model as being consistent with LDD convection. We also present a Jupiter model with $l_H$ decreasing by 10 per cent within 1 and 3 Mbar and a toy composition gradient $c_3=\beta\nabla_{\mu}(\alpha_{\mu}/\alpha_T)\equiv 0.1$, implying $\beta < 1$. The resulting $\Rinv$ values of the latter model are indeed fully consistent with LDD convection\footnote{That model is actually based on a slightly different 
modification of the H/He diagram, say modified-2b, as can be seen from the onset of demixing at 1 instead of 1.4 Mbar, 
but this has negligible effect on the following results and discussion.}.
We proceed with both models (the \emph{cyan} $\beta=1$ model and the \emph{red} $c_3$ model) to compute the cooling times.

\subsubsection{Cooling time}\label{sec:evol}

\begin{figure}
\includegraphics[width=0.45\textwidth]{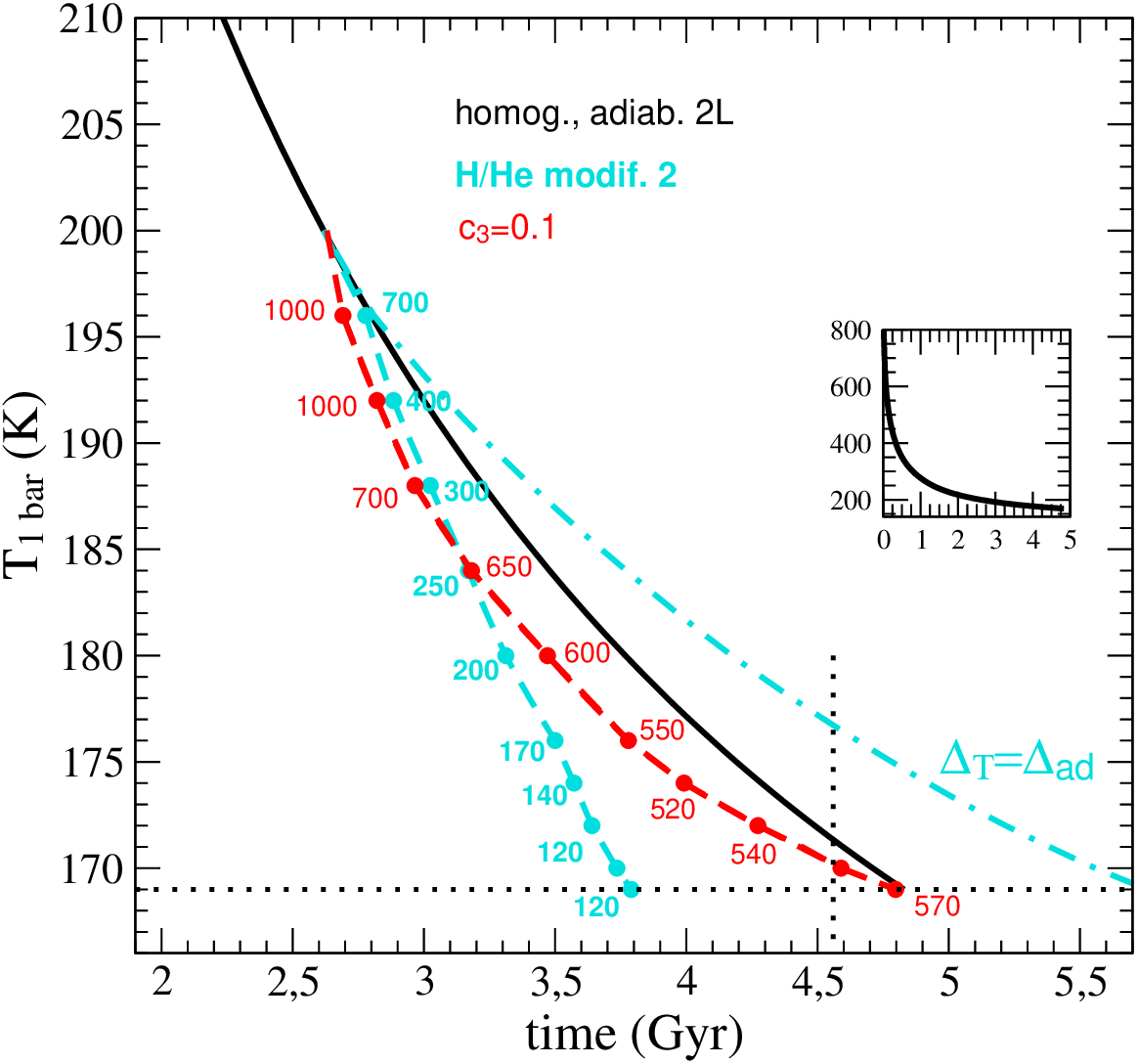}
\caption{\label{fig:evol2}Time evolution of Jupiter with He rain and LDD convection using the modified-2 H/He phase diagram (\emph{cyan with circles}) in comparison to adiabatic homogeneous evolution without He rain (\emph{black}),  adiabatic inhomogeneous evolution with He rain (\emph{cyan, dot-dashed}), and for He rain with LDD convection but setting $c_3=0.1$ in Eq. \ref{eq:c3} (\emph{grey, dashed}). Numbers give $l_H$ in m. For the evolution the Fortney model atmosphere is applied.}
\end{figure}

In Figure \ref{fig:evol2} we present the cooling times for our two modified-2 H/He phase diagram based Jupiter models with LDD convection and adjusted layer heights over time to yield the low $\Rinv$ values as described above.

It is an important result of this work that our only model that we consider consistent with the $\Rinv$-range of LDD convection, being based on a (modified) first-principles derived H/He phase diagram, our fundamental assumptions, and a state-of-the-art Jupiter model atmosphere \citep{Fortney+11} has a cooling time of only 3.8~Gyr. 

With a cooling time of only 3.8 Gyr, our thermal evolution model appears to indicate room for additional complexities in our understanding of Jupiter's structure and evolution.  Standard quasi-homogeneous, adiabatic  models that reproduce all observational constraints already result in a cooling time in good agreement with Jupiter's known age \citep{N+12}. 
However, those models with (sharp) gradients in the abundance of helium and heavy elements as constructed by  \citet{N+12} and, more recently, \citet{Becker+14} ignore the heat transport and temperature gradient in the layer boundary zone(s) altogether and thus are \emph{physically inconsistent}. In this paper we have instead tried for more self-consistency, but at the expense of sacrificing the previous apparent understanding of Jupiter's thermal evolution. We note that the only suite of physically self-consistent Jupiter models are the two-layer models by \citet{Militzer+08}; however, they do not reproduce the gravity field data, neither do they provide an explanation for Jupiter's atmospheric He depletion. 

We like to emphasize the importance of including a theory for the heat transport in the computation of inhomogenous thermal evolution. Omitting such a theory by assuming an adiabatic temperature profile ($\nabla_T=\nabla_{ad}$) would yield a signigicant cooling time prolongation, see the \emph{cyan dot-dashed} curve in Figure \ref{fig:evol2}. Such a model may be applicable to Saturn, where He droplets may rain down to the core \citep{FH03, Puestow+14}.
Our explorations show, however, that the application of such a theory (here: semi-convection) in combination with a H/He phase diagram does not necessarily directly lead to a balance between the additional gravitational energy and the energy transported through the inhomogeneous zone, that yields the correct cooling time. Moreover, our explorations rule out the case $\beta=0$, which would imply an adiabatic, convective, inhomogeneous interior that cools too slowly.

On the other hand, relaxation of the fundamental assumption ($iii$) is represented by the model with  toy composition gradient $c_3=0.1$. Because of the lower composition gradient ($\beta<1$), it requires lower superadiabaticities to meet the $\Rinv$-constraint (Figures \ref{fig:Rparam}, \ref{fig:fluxes}). Accidentally, in this case the desired energy-balance is exactly reached, see the \emph{red dashed} curves in Figures \ref{fig:evol2} and \ref{fig:TcRpss2}: the deep internal entropy decreases a bit slower in time than in the adiabatic homogeneous case but the outward heat flux is maintained by the higher internal temperatures, so that the cooling time remains unchanged compared to the homogeneous, adiabatic case without He rain. We evaluate the $c_3$ model to be our best-case Jupiter model, as it satisfies all constraints and would even be consistent with a magnetic dynamo in the convective interior below the demixing zone. This  model has internal layer heights of 500-1000 m in Jupiter, slowly decreasing in time, corresponding to  $\sim$20,000 layers in Jupiter's current He rain zone. Comparing $c_3=0.1$ and the $c_3$-profile for $\beta=1$ as shown in Figure \ref{fig:fluxes}, and considering $\Rinv=1$--3, we derive $\beta=0.1/(0.45\mbox{--}0.9) \approx$ 0.05--0.25, in agreement with our estimate from Section \ref{sec:res1LDD}.

\begin{figure}
\includegraphics[width=0.45\textwidth]{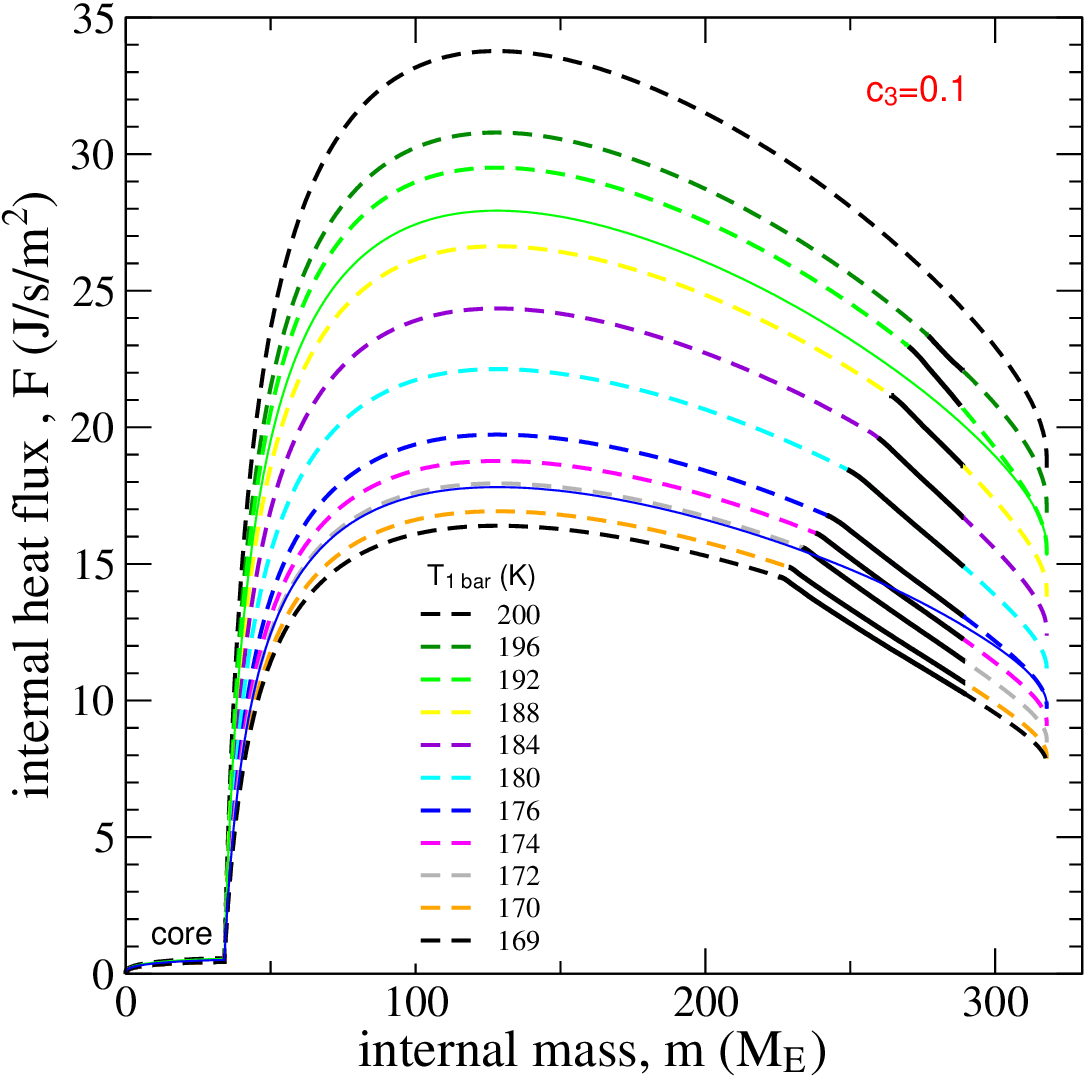}
\caption{\label{fig:flux_c3}Internal heat flux profiles of our best-case Jupiter model with the toy composition gradient $c_3=0.1$. \emph{Coloures} are for different {\Tsurf} values during the evolution. \emph{Black solid parts} along the \emph{dashed curves} indicate the He rain region, while \emph{thin solid} curves are for homogeneous, adiabatic evolution.}
\end{figure}

Figure \ref{fig:flux_c3} shows the internal heat flux profiles of our best-case Jupiter models. Due to the additional energy from the He rain, the internal flux is \emph{higher} than in the absence of He rain (\emph{colored solid}) while the heat flux drops in the semi-convective rain zone. It expands over time.

\section{Discussion and Outlook}\label{sec:discuss}

\subsection{Comparison with the theory of L \& C (2012)}\label{sec:discussLC12}

The present work shares similarities with the work of LC12. Both groups derive an expression for the total heat flux $F$ as a function of the superadiabaticity and the layer height which allows one to pick that $\nabla_{T}$ value, for a given $l_H$ value, that results into the desired flux value. A difference lies in the Nu$_T$--$Ra_{\star}$ relations used to calculate $F$. For the Nu$_T$--$Ra_{\star}$ relation we use a fit formula (Eqs.~\ref{eq:FT2}, \ref{eq:NuTansatz}) to the heat flux ''measurements'' from hydrodynamic simulations, where the heat flow $F_{\rm LDD}$ through a small (2--3) number of alternating layers and interfaces is determined through the imposed average vertical temperature and density gradients. As found in the simulations, the heat flow through the layers is reduced compared to the vigorous convective case ($F_{\rm LDD}<F_{\rm conv}$), while the heat flow through the interfaces is enhanced compared to the purely diffusive case 
($F_{\rm LDD} > F_{\rm cond}$) for a given temperature gradient. The latter result is also reflected in our Jupiter model with LDD convection, see the lower panel of Figure \ref{fig:fluxes}.

In contrast, LC12 derive separate analytic expressions for the heat flows in the convective layers and the diffusive interfaces, respectively. For $F_{\rm conv}$ they use a generalized mixing length expression, where the layer height would correspond to the mixing length. In particular, their generalization resembles the fit formula (\ref{eq:NuTansatz}) for $\rm Pr\,Ra\gg 1$, $f_T=1$, and $0.2 \leq a \leq 0.5$. The latter degree of freedom causes a wider range of their solutions, while we use $a=1/3$. For the diffusive heat flux they use the standard expression (\ref{eq:Fcond}). They invert these expressions separately to obtain the local temperature gradient in the layers ($\nabla_T$) and in the interfaces ($\nabla_d$) that yield the same total flux. They then build the average temperature gradient across many layers by weighting the two local temperature gradients with the respective size of the layers ($l_H$) and of the interfaces ($\delta_T$), where $l_H$ is considered a free parameter to be narrowed down by constraints from Jupiter structure and evolution modeling, while $\delta_T$ is derived from the assumption of equal thermal diffusive and convective time-scales. 

Most interestingly, \citet{LC13} see the  opposite response of the cooling time (Saturn's) to both the presence of LDD convection and to the size of the layer height. They find a \emph{prolongation} of the cooling time, that even \emph{increases} with decreasing layer height. We argue that the solution to that apparent discrepancy is related to the ``crossover'' of their (Saturn) cooling curves over time, meaning that the models with composition gradients eventually must have higher luminosities than homogeneous models at old ages, even though they are less luminous at young ages.  
We ran test calculations assuming a central super-adiabatic region from young ages and ideed observed a slight cooling time prolongation. In our Jupiter models with He rain only, the crossover point has not been reached yet.

\subsection{Other planets: Saturn}

Saturn is the canonical example of a planet where H/He demixing is suggested to occur; in this case to explain its high luminosity. Unfortunately, Saturn's atmospheric helium abundance is not well known, with measurement mass fractions ranging from 1--11 per cent from the Voyager radio occultation and Voyager infrared spectroscopy experiments to 18--25 per cent from their later re-analyses \citep{Conrath84,ConGau00}, see also Figure \ref{fig:Yatm_T1}.  Assuming $\Yatm=0.20$ for present Saturn, \citet{FH03} could explain the observed excess luminosity by a modified Pfaffenzeller et al.~H/He phase diagram, which predicts He-rain down to the core. 
Due to the current uncertainties in Saturn's $\Yatm$ value and in the H/He phase diagrams, an inhomogeneous region and LDD convection as a result of He rain can, however, not be excluded for Saturn, in particular as a transitional state prior to He-layer formation. We here emphasize the importance of an accurate measurement of Saturn's He abundance, most accurately done by sending an entry probe \citep{Fortney+09,Mousis13}.

\subsection{Other planets: Uranus}

Uranus is the canonical example of a planet where inhibited heat transport is suggested to occur, in this case to explain its low luminosity \citep{Hubbard+95}. Stable stratification, whether in the form of semi-convection or diffusion, has not been taken into account explicitly yet in Uranus structure and evolution models. Our results suggest that if stable stratification occurs in Uranus, then maybe not in the form of semi-convection, because the $\Rinv$-range where semi-convection can operate is already small for Jupiter (1-- $\approx$10) and may even reduce to 1--2 as a result of the large Prantl number (Pr $>1$) of water. Stable stratification with suppression of heat flux from the deep interior may be more likely realized in Uranus than in Jupiter because of the planets' difference in total mass by a factor of 20.

\subsection{Stars}

The difficulties we face in developing a semi-convective zone model for Jupiter is somewhat 
at odds given the long history in the treatment of  semi-convection in stars, \citep[e.g.][]{Stev79,Langer85}. 
For instance, in massive stars of 15--30 solar masses, a semi-convective zone is suggested to form 
between the He-burning convective core and the overlying H-rich envelope. In this case, 
a composition gradient arises from the formation of heavier C/O in the central core.
Several relevant differences can be stated that impede the adoption of a of well-studied scheme to planets: 
First, stars have nuclear reactions as a dominant internal heat source while giant planets get their luminosity 
mostly from the slow cooling of the ions so that the internal luminosity must collapse if the planet's 
deep interior is prevented from cooling, which may explain part of our difficulties in finding a solution
for the ODD case with the modified-2 H/He data; 
Second, in stars the radiative gradient is of the order of the adiabatic gradient \citep{Gabriel14} 
and thus the actual temperature gradient is often \citep{Gabriel14,Vazan14} but not always \citep{Stev79,Langer85,DingLi14} 
equaled with either one, whereas in planets $\nabla_{rad}\gg\nabla_{ad}$ and thus 
$\nabla_{rad}\gg\nabla_{T}$ unless the internal heat flux is suppressed. We here applied a description of DD 
convection to obtain an estimate on  $\nabla_{T}$. A different theory could be applied as well.
Third, semi-convection in massive stars can greatly alter the distribution of elements through enhanced diffusion 
\citep{Langer85} despite $\tau \ll 1$ in stars, while here we assume that a composition 
gradient is permanently maintained through steady demixing, 
and that diffusion plays no role despite of $\tau < 1$ only. In particular --and here we come back
to the fundamental caveats of this work-- the combination of $\tau < 1$ and of diffusive transport along 
the composition gradient together built the essence of the theory of semi-convection. Here we have assumed that the (downward) transport of He through sedimentation and (upward) transport through diffusion
or convection happen linearly superimposed so that the essence of semi-convection does not get undermined.
The effect of diffusive transport on the helium distribution in a cooling giant planet remains to be investigated.

\subsection{Adiabatic, homogeneous models}

Adiabatic, homogeneous evolution models yield good, or perhaps slightly too long cooling ages for Jupiter \citep{SG04,Fortney+11,N+12}. Those previous models ignore the finite gradients in composition and temperature between a He-poor outer and an He-rich inner envelope, and/or between an heavy element-poor to an heavy-element rich deep interior by assuming ad-hoc layer boundaries with infinite gradients in composition and entropy \citep{Guillot+97,GZ99,SG04,N+08,N+12}. Such a treatment of the heavy elements has raised suspicion about its physical justification and led \citet{SG04} to  assume a homogeneous distribution of heavy elements (which restricts the number of H/He EOS that can cope with the reduced degrees of freedom). \citet{Militzer+08} dropped both the discontinuity in He and in heavy elements  by assuming a fully adiabatic, homogeneous envelope (although that model was inconsistent with the observed gravity field).
Here we have taken a first step to move beyond the successful but ad-hoc picture of Jupiter by treating finite helium and temperature gradients, albeit in a still vastly simplified and perheps premature manner.

\subsection{Z-distribution}

We ignored the distribution of heavy elements in Jupiter's mantle. 
On the other hand, \citet{LC13} have solely considered the $Z$-distribution (for Saturn), with the result of a cooling time prolongation of several Gyrs (for Saturn). This suggests that perhaps for Jupiter we have only tackled part of the problem, and both distributions (He and $Z$) need to be treated simultaneously. According to the models by \citet{LC12,LC13}, Jupiter's interior could already be super-adiabatic when He-rain starts to operate. This would reduce our need to modify the H/He phase diagram in the $\beta=1$ case. 
It remains to be investigated how large an inhomogeneous region could be while still allowing for the generation of a magnetic field.

\subsection{The $\beta \ll 1$ case}

Our  model with an enforced lower composition gradient ($c_3=0.1$) is the only one to satisfy all considered constraints. A lower composition gradient ($\beta<1$) than imposed by the H/He phase diagram alone may have a variety of different origins, like contributions from He rain-out from rising parcels, from the temperature- and composition-dependence of the metallization of hydrogen including dissociation and ionization, and from a change in heavy element abundance in the stratified fluid background.
It is also possible that He-rain requires some super-cooling before the He-droplets can form and fall. 
Currently, these contributions are essentially unknown. A lower composition gradient ($\beta\ll 1$) could also help to stop the run-away effect between gradients in $T$ and in $Y$ \citep{FH03}, so that a solution with a fully convective while still superadiabatic demixing zone may become possible.

\subsection{The Earth as a guide}\label{sec:Earth}

Droplet formation and rain in the Earth's atmosphere are extensively studied processes. We review basic properties and discuss possible implications for He rain in a giant planet.

It is well-known from daily experience that rain does not fall from a clear sky. In fact, the formation of clouds or hazes
proceeds that of rain. In our study we have neglected clouds/hazes. Rain-less clouds are accumulations of microscopic ($\sim 20 \mu$m) water droplets of insufficient size for rain-out. Those droplets form if the partial pressure of dissolved water in a vertically moving air parcel exceeds the water condensation pressure or, if the air parcel is cold enough, the sublimation pressure. 

In the Earth's atmosphere, the conditions for droplet formation depend on the water phase diagram, on the abundance of water, on the background $P$--$T$ profile, also called the environmental lapse rate, and on nucleation seeds.  
The phase diagram of water and the applied one of H/He share similarities such as phase boundaries $T(P)$ that increase with pressure and an enhanced solubility with increasing temperature, suggesting the Earth's atmosphere to be a reasonable guide for understanding He rain in giant planets. On the other hand, obvious differences exist as well.
In the Earth's atmosphere, the background profile is considered to be given. It can have a number of origins that to first order do not depend on the vertical distribution of water, such as horizontal winds, surface heating, or particulate pollution.   Contrary, in the He rain case the distribution of helium might determine the background temperature profile to first order through the inhibition of large-scale convection as suggested in this work. This is because helium in a H/He planet is a major constituent, while water in the Earth's N$_2$/O$_2$ atmosphere is not. Furthermore, the background He abundance is assumed to follow the phase diagram \citep{SS77b} while the Earth-atmospheric water abundance is generally under-saturated.

On Earth, cloud formation and eventually rain require vertical motions in the atmosphere (a prominent exception being fog). Through vertical motions, upwelling air parcels can expand adiabatically. As long as humidity, i.e. the ratio between the actual vapor pressure and the saturation pressure, is below 100 per cent, the parcels' $T$-$P$ profile follows a \emph{dry} adiabat. Conversely, the \emph{wet} adiabat of air is characterized by 100 per cent humidity  so that the parcels' water abundance follows the phase diagram while the excess water condenses out and is dispensed to the environment, in form of microscopic droplets. Due to latent heat release upon condensation, the wet adiabat is flatter than the dry one. 

In this work we have for two reasons assumed no rain-out from rising parcels ($\beta=1$) corresponding to a super-saturated wet adiabat. First, this fundamental assumption was inherited from mixing length theory where the concentration of blobs is supposed to remain conserved because of low particle diffusivities. However, the Earth's atmosphere tells us  that loss of particles from a moving parcel can well happen ($\beta<1$) if the underlying process is non-diffusive in nature. Second, the computation of a wet adiabat requires knowledge of the latent heat, while the latent heat from He droplet formation is unknown at present. Concluding, although our assumption of $\beta=1$ can be deemed inappropriate, both from our modeling results and the known properties of wet water adiabats, the direct computation of the latter one for a demixing H/He mixture is subject to great uncertainties at present. 

In the Earth's atmosphere, larger droplets can sink faster, thereby colliding with smaller ones. Coalescence then leads to droplet growth and eventually to rain. In addition to this basic hydrodynamic-gravitational process, background turbulent fluctuations can have a major impact on the rate and efficiency of collisions and thus on the initiation of precipitation \citep{Tisler05,Wang05}. Rainfall on Earth is also strongly affected by aerosols in a number of ways \citep{Ganguly12}.

Such gross properties may apply also to He rain in a H/He giant planet polluted with heavy elements. 
However, at present it is unknown how ---and if at all--- demixing, cloud formation, and rainfall fit into the picture of LDD convection. One might speculate that interfaces occur as a result of He cloud/He haze formation, and that He rainfall occurs like Virga on Earth. It could also be possible that the formation of sufficiently large He rain droplets from He clouds takes a long time so that the assumption of instantaneous sedimentation becomes irrelevant. In that case, convection may persist in the presence of cumulus-like clouds. A convective demixing zone is not excluded if $\beta \ll 1$.

At the very least, this work demonstrates how little we know about the interior of giant planets that are supposed to be ``simple,'' and that an enormous interdisciplinary effort might be necessary to gain reliable insight and, as discussed in the Introduction, provide the missing ``fourth leg'' that could support a complete theory. Such an effort should include numerical simulations.

\subsection{Numerical simulations and prospects}

Numerical simulations have helped to identify the demixing phase diagram of H/He (e.g. Lorenzen 2011), 
to quantify the flux of heat and solute through semi-convective layers (\citealp[e.g.][]{Wood13}), 
and to predict the formation of water clouds and rainfall (e.g.~\citealp{Ganguly12}).
While each of them are subject to uncertainties on their own (H/He: the function $T_{dmx}(P,x_{\rm He})$;
LDD: layer merging is seen but with unknown convergent state; rain: the role of nucleation seeds), 
a combined effort of these three branches at least might be necessary to advance our understanding
of the processes at work in a Jovian planet. Future numerical simulations should include clouds/hazes 
and the time-scale for raindrop formation in a mixture where the solute is a major constituent; 
studies of (DD) convection with a solute that can move by non-diffusive processes, and simulations 
of the formation and growth of microscopic droplets in a saturated H/He mixture. Simulations should
also addresss the behavior of heavy elements in the presence of H/He demixing, as it has been done 
for neon \citep{WilMil10}.

\section{Summary}\label{sec:summ}

Our results clearly show that the development of a self-consistent Jupiter model is a complex enterprise.
We have presented new Jupiter evolution models that we think are an advance over previous work in that they include helium rain as indicated by Jupiter's He depleted atmosphere, as well as double diffusive convection as a result of the composition gradient in the rain region. Our goal, started with this paper, is to develop thermal evolution and internal structure models of giant planets that include such properties in a self-consistent manner, which we think could drive a new era of planetary modeling, well timed with new data from \emph{Juno}, for Jupiter, and \emph{Cassini}, for Saturn.  An equivalent approach for understanding composition gradients in low-mass objects such as Ganymede and Mercury is in progress, where the origin of the observed magnetic field is coupled to stable stratification in an iron core as a result of Fe-snow \citep{Rueck14g}.

This paper is dedicated to a thorough description of our applied method to iteratively solve for self-consistent profiles of composition and temperature, given a H/He phase diagram and a prescription for the corresponding superadiabaticity.  To determine  the He profile we used the \citet{Lorenzen09,Lorenzen11} phase diagram, and modified it in order to obtain an atmospheric He abundance in agreement with the \emph{Galileo} probe measurement. To determine the temperature profile we used a description of semi-convection, either in the form of layered or of oscillatory double diffusive convection, which was adapted from numerical simulations by \citet{Mirouh12} and \citet{Wood13}. Furthermore, we applied the SCvH EOS, as it conveniently provides the entropy and partial derivatives. 

We presented a wide range of models that we successively ruled out because of lack of self-consistency. Our main results are that (i) adiabatic models with He rain lead to a $\sim 0.7$ Gyr too long cooling time;  (ii) ODD convective models are difficult to reconcile with Jupiter's observed high heat flux, (iii) LDD convective models with prohibited particle exchange between convective eddies and the ambient fluid ($\beta=1$) lead to a shortfall of the cooling time by 0.5--1 Gyr relative to the age of the Solar system, (iv) LDD convective models with allowed (but not explicitly modeled) loss of He droplets from upwelling eddies to the ambient fluid ($\beta<1$) satisfy all constraints and yield the proper cooling time. Those  models give a He rain zone roughly between 1 and 3--4.5 Mbars, depending on the H/He phase diagram, and a subdivision into 10,--20,000 layers ($l_H\approx 100$--1000 m). The modest superadiabaticity of $\sim 0.1$ enhances the core temperature by a $\sim 1000$ K only which we expect to slightly enhance the inferred core mass.
(v) A superadiabatic convective demixing zone is not excluded provided $\beta\ll 1$.
Our results suggest that our understanding of the transport of energy and particles through a zone where H/He phase separation occurs still needs considerable work, and that the last word about the H/He phase diagram itself may not have been said yet. 
Additional physical processes likely effect Jupiter's cooling, including heavy element gradients, compositional exchange between displaced mass elements and their surroundings, and He cloud formation, none of which are yet well accommodated for in models of semi-convection and H/He phase separation.

In closing we note that in superadiabatic regions the Brunt-V\"ais\"ala frequency is non-zero so that gravity waves can be excited. We are hopeful that seismological observations, challenging as they may be, will one day resolve ambiguities in Jupiter's (and Saturn's) internal structure.

\section*{Acknowledgments}

The anonymous referee is acknowledged for an insightful report which improved the discussion of the presented work.  
We thank Sebastien Hamel, Fran\c{c}ois Soubiran, Johannes Wicht, Tina R{\"u}ckriemen, Ronald Redmer, and Winfried Lorenzen for fruitful discussions, and Jeremy Leconte for hints on the internal heat flux profile.
This work was supported by NSF grant AST-1010017.

\bibliographystyle{mn2e}
\bibliography{./ms-refs}

\begin{thebibliography}{}

\bibitem[\protect\citeauthoryear{Becker, Lorenzen, Fortney, Nettelmann,
  Sch{\"o}ttler \& Redmer}{Becker et~al.}{2014}]{Becker+14}
Becker A.,  Lorenzen W.,  Fortney J.~J.,  Nettelmann N.,  Sch{\"o}ttler M.,
  Redmer R.,  2014, ApJ, accepted

\bibitem[\protect\citeauthoryear{Becker, Nettelmann, Holst \& Redmer}{Becker
  et~al.}{2013}]{Becker13}
Becker A.,  Nettelmann N.,  Holst B.,    Redmer R.,  2013, PRB, 88, 045122

\bibitem[\protect\citeauthoryear{Chabrier \& Baraffe}{Chabrier \&
  Baraffe}{2007}]{CB07}
Chabrier G.,  Baraffe I.,  2007, ApJ, 661, L81

\bibitem[\protect\citeauthoryear{Chabrier, Saumon, Hubbard \& Lunine}{Chabrier
  et~al.}{1992}]{Chabrier+92}
Chabrier G.,  Saumon D.,  Hubbard W.,    Lunine J.,  1992, ApJ, 391, 826

\bibitem[\protect\citeauthoryear{Conrath \& Gautier}{Conrath \&
  Gautier}{2000}]{ConGau00}
Conrath B.,  Gautier D.,  2000, Icarus, 144, 124

\bibitem[\protect\citeauthoryear{Conrath, Gautier, Hanel \& Hornstein}{Conrath
  et~al.}{1984}]{Conrath84}
Conrath B.~J.,  Gautier D.,  Hanel R.~A.,    Hornstein J.~S.,  1984, Astrophys.
  J, 282, 807

\bibitem[\protect\citeauthoryear{Ding \& Li}{Ding \& Li}{2014}]{DingLi14}
Ding C.~Y.,  Li Y.,  2014, MNRAS, 438, 1137

\bibitem[\protect\citeauthoryear{Fortney \& Hubbard}{Fortney \&
  Hubbard}{2003}]{FH03}
Fortney J.~J.,  Hubbard W.~B.,  2003, Icarus, 164, 228

\bibitem[\protect\citeauthoryear{Fortney, Ikoma, Nettelmann, Guillot \&
  Marley}{Fortney et~al.}{2011}]{Fortney+11}
Fortney J.~J.,  Ikoma M.,  Nettelmann N.,  Guillot T.,    Marley M.~S.,  2011,
  ApJ, 729, 32

\bibitem[\protect\citeauthoryear{Fortney, Zahnle, Baraffe \& Burrows}{Fortney
  et~al.}{2009}]{Fortney+09}
Fortney J.~J.,  Zahnle K.,  Baraffe I.,    Burrows A. e.~a.,  2009, eprint
  arxiv:0911:3699

\bibitem[\protect\citeauthoryear{French, Becker, Lorenzen, Nettelmann,
  Bethkenhagen, Wicht \& Redmer}{French et~al.}{2012}]{French12}
French M.,  Becker A.,  Lorenzen W.,  Nettelmann N.,  Bethkenhagen M.,  Wicht
  J.,    Redmer R.,  2012, ApJS, 202, A5

\bibitem[\protect\citeauthoryear{Gabriel, Noels, Montalban \& Miglio}{Gabriel
  et~al.}{2014}]{Gabriel14}
Gabriel M.,  Noels A.,  Montalban J.,    Miglio A.,  2014, A \& A, 569, A63

\bibitem[\protect\citeauthoryear{Ganguly, Rasch, Wang \& Yoon}{Ganguly
  et~al.}{2012}]{Ganguly12}
Ganguly D.,  Rasch P.,  Wang H.,    Yoon J.-H.,  2012, 2012, 117, D13209

\bibitem[\protect\citeauthoryear{Gautier, Conrath, Flasar, Hanel, Kunde, Chedin
  \& Scott}{Gautier et~al.}{1981}]{Gautier81}
Gautier D.,  Conrath B.,  Flasar M.,  Hanel R.,  Kunde V.,  Chedin A.,    Scott
  N.,  1981, J. Geophys. Res., 86, 8713

\bibitem[\protect\citeauthoryear{Graboske, Olness, Pollack \&
  Grossman}{Graboske et~al.}{1975}]{Graboske75}
Graboske H.~C.,  Olness R.~J.,  Pollack J.~B.,    Grossman A.~S.,  1975, ApJ,
  199, 265

\bibitem[\protect\citeauthoryear{Gudkova \& Zharkov}{Gudkova \&
  Zharkov}{1999}]{GZ99}
Gudkova T.,  Zharkov V.~N.,  1999, Planet. Space Sci., 47, 1201

\bibitem[\protect\citeauthoryear{Guillot, Gautier \& Hubbard}{Guillot
  et~al.}{1997}]{Guillot+97}
Guillot T.,  Gautier D.,    Hubbard W.~B.,  1997, Icarus, 130, 534

\bibitem[\protect\citeauthoryear{Guillot, Stevenson, Hubbard \& Saumon}{Guillot
  et~al.}{2004}]{Guillot03}
Guillot T.,  Stevenson D.~J.,  Hubbard W.~B.,    Saumon D.,  2004, {The
  interior of Jupiter}.
pp 35--57

\bibitem[\protect\citeauthoryear{Hubbard, Guillot, Marley, Burrows, Lunine \&
  Saumon}{Hubbard et~al.}{1999}]{Hubbard99}
Hubbard W.~B.,  Guillot T.,  Marley M.~S.,  Burrows A.,  Lunine J.~I.,
  Saumon D.~S.,  1999, Planet.~Space Sci., 47, 1175

\bibitem[\protect\citeauthoryear{Hubbard, Podolak \& Stevenson}{Hubbard
  et~al.}{1995}]{Hubbard+95}
Hubbard W.~B.,  Podolak M.,    Stevenson D.~J.,  1995, in Cruishank ed., ,
  Neptune and Triton.
University of Arizona, Tucson, p.~109

\bibitem[\protect\citeauthoryear{Klepeis, Schafer, Barbee~III \& Ross}{Klepeis
  et~al.}{1991}]{Klepeis91}
Klepeis J.~E.,  Schafer K.~J.,  Barbee~III T.,    Ross M.,  1991, Science, 254,
  986

\bibitem[\protect\citeauthoryear{Langer, El~Eid \& Fricke}{Langer
  et~al.}{1985}]{Langer85}
Langer N.,  El~Eid M.,    Fricke K.~J.,  1985, A \& A, 145, 179

\bibitem[\protect\citeauthoryear{Leconte \& Chabrier}{Leconte \&
  Chabrier}{2012}]{LC12}
Leconte J.,  Chabrier G.,  2012, A\&A, 540, A20

\bibitem[\protect\citeauthoryear{Leconte \& Chabrier}{Leconte \&
  Chabrier}{2013}]{LC13}
Leconte J.,  Chabrier G.,  2013, Nature Geoscience, 6, 347

\bibitem[\protect\citeauthoryear{Lorenzen, Holst \& Redmer}{Lorenzen
  et~al.}{2009}]{Lorenzen09}
Lorenzen W.,  Holst B.,    Redmer R.,  2009, Phys. Rev. Lett., 102, 5701

\bibitem[\protect\citeauthoryear{Lorenzen, Holst \& Redmer}{Lorenzen
  et~al.}{2011}]{Lorenzen11}
Lorenzen W.,  Holst B.,    Redmer R.,  2011, Physical Review B, 84, 235109

\bibitem[\protect\citeauthoryear{Loubeyre, LeTullec \& Pinceaux}{Loubeyre
  et~al.}{1985}]{Loubeyre85}
Loubeyre P.,  LeTullec R.,    Pinceaux J.-P.,  1985, PRB, 32, 7611

\bibitem[\protect\citeauthoryear{Militzer, Hubbard, Vorberger, Tamblyn \&
  Bonev}{Militzer et~al.}{2008}]{Militzer+08}
Militzer B.,  Hubbard W.~B.,  Vorberger J.,  Tamblyn I.,    Bonev S.~B.,  2008,
  ApJ, 688, L54

\bibitem[\protect\citeauthoryear{Mirouh, Garaud, Stellmach, Traxler \&
  Wood}{Mirouh et~al.}{2012}]{Mirouh12}
Mirouh G.~M.,  Garaud P.,  Stellmach S.,  Traxler A.~L.,    Wood T.~S.,  2012,
  ApJ, 750, 61

\bibitem[\protect\citeauthoryear{Morales, McMahon, Pierleoni \&
  Ceperley}{Morales et~al.}{2013}]{Morales13b}
Morales M.,  McMahon J.~M.,  Pierleoni C.,    Ceperley D.~M.,  2013, PRL, 110,
  065702

\bibitem[\protect\citeauthoryear{Morales, Hamel, Caspersen \&
  Schwegler}{Morales et~al.}{2013}]{Morales13}
Morales M.~A.,  Hamel S.,  Caspersen K.,    Schwegler E.,  2013, PRB, 87,
  174105

\bibitem[\protect\citeauthoryear{Morales, Schwegler, Ceperley, Pierleoni, Hamel
  \& Caspersen}{Morales et~al.}{2009}]{Morales09}
Morales M.~A.,  Schwegler E.,  Ceperley D.,  Pierleoni C.,  Hamel S.,
  Caspersen K.,  2009, PNAS, 106, 1324

\bibitem[\protect\citeauthoryear{Mousis}{Mousis}{2013}]{Mousis13}
Mousis O.,  2013, Pl.Sp.Sci., submitted

\bibitem[\protect\citeauthoryear{Nettelmann, Becker, Holst \&
  Redmer}{Nettelmann et~al.}{2012}]{N+12}
Nettelmann N.,  Becker A.,  Holst B.,    Redmer R.,  2012, ApJ, 750, A52

\bibitem[\protect\citeauthoryear{Nettelmann, Holst, Kietzmann, French, Redmer
  \& Blaschke}{Nettelmann et~al.}{2008}]{N+08}
Nettelmann N.,  Holst B.,  Kietzmann A.,  French M.,  Redmer R.,    Blaschke
  D.,  2008, ApJ, 683, 1217

\bibitem[\protect\citeauthoryear{Niemann, Atreya \& Carignan}{Niemann
  et~al.}{1998}]{Niemann98}
Niemann H.,  Atreya S.~K.,    Carignan G. R. e.~a.,  1998, JGR, 103, 22831

\bibitem[\protect\citeauthoryear{Orton \& Ingersoll}{Orton \&
  Ingersoll}{1976}]{OrtonIng76}
Orton G.,  Ingersoll A.~P.,  1976, in IAU Colloq. 30: Jupiter: Studies of the
  Interior, Atmosp here, Magnetosphere and Satellites U Arizona Press, p.~206

\bibitem[\protect\citeauthoryear{Pfaffenzeller, Hohl \& Ballone}{Pfaffenzeller
  et~al.}{1995}]{Pfaffenzeller96}
Pfaffenzeller O.,  Hohl D.,    Ballone P.,  1995, PRL, 74, 2599

\bibitem[\protect\citeauthoryear{P{\"u}stow, Nettelmann, Lorenzen \&
  Redmer}{P{\"u}stow et~al.}{2014}]{Puestow+14}
P{\"u}stow R.,  Nettelmann N.,  Lorenzen W.,    Redmer R.,  2014, in prep.

\bibitem[\protect\citeauthoryear{Radko}{Radko}{2003}]{Radko03}
Radko T.,  2003, J.~Fluid Mech., 497, 365

\bibitem[\protect\citeauthoryear{Rosenblum, Garaud, Traxler \&
  Stellmach}{Rosenblum et~al.}{2011}]{Rosenblum11}
Rosenblum E.,  Garaud P.,  Traxler A.,    Stellmach S.,  2011, ApJ, 731, 61

\bibitem[\protect\citeauthoryear{R\"uckriemen, Breuer \& Spohn}{R\"uckriemen
  et~al.}{2014}]{Rueck14g}
R\"uckriemen T.,  Breuer H.,    Spohn T.,  2014, LPI, 45, 2454

\bibitem[\protect\citeauthoryear{Salpeter}{Salpeter}{1973}]{Salpeter73}
Salpeter E.~E.,  1973, ApJ, 181, L83

\bibitem[\protect\citeauthoryear{Saumon, Chabrier \& "van Horn"}{Saumon
  et~al.}{1995}]{SCvH95}
Saumon D.,  Chabrier G.,    "van Horn" H.~M.,  1995, ApJ, 99, 713

\bibitem[\protect\citeauthoryear{Saumon \& Guillot}{Saumon \&
  Guillot}{2004}]{SG04}
Saumon D.,  Guillot T.,  2004, ApJ, 609, 1170

\bibitem[\protect\citeauthoryear{Saumon, Hubbard, Chabrier \& van Horn}{Saumon
  et~al.}{1992}]{Saumon+92}
Saumon D.,  Hubbard W.~B.,  Chabrier G.,    van Horn H.~M.,  1992, ApJ, 391,
  827

\bibitem[\protect\citeauthoryear{Soderlund, Heimpel, King \& Aurnou}{Soderlund
  et~al.}{2013}]{Soderlund13}
Soderlund K.~M.,  Heimpel M.~H.,  King E.~M.,    Aurnou J.~M.,  2013, Icarus,
  224, 97

\bibitem[\protect\citeauthoryear{Soubiran, Mazevet, Winisdoerffer \&
  Chabrier}{Soubiran et~al.}{2013}]{Soubiran13}
Soubiran F.,  Mazevet M.,  Winisdoerffer C.,    Chabrier G.,  2013, Phys. Rev.
  B, 87, 165114

\bibitem[\protect\citeauthoryear{Stevenson}{Stevenson}{1975}]{Stevenson75}
Stevenson D.~J.,  1975, PRB, 12, 3999

\bibitem[\protect\citeauthoryear{Stevenson}{Stevenson}{1979}]{Stev79}
Stevenson D.~J.,  1979, MNRAS, 187, 129

\bibitem[\protect\citeauthoryear{Stevenson}{Stevenson}{1998}]{Stev98}
Stevenson D.~J.,  1998, J. Phys.: Condens. Matter, 10, 11227

\bibitem[\protect\citeauthoryear{Stevenson \& Salpeter}{Stevenson \&
  Salpeter}{1977}]{SS77b}
Stevenson D.~J.,  Salpeter E.,  1977, ApJS, 35, 239

\bibitem[\protect\citeauthoryear{Tisler, Zapadinsky \& Kulmala}{Tisler
  et~al.}{2005}]{Tisler05}
Tisler P.,  Zapadinsky E.,    Kulmala M.,  2005, Geo. Res. Lett., 32, L06806

\bibitem[\protect\citeauthoryear{Vazan, Helled, Kovetz \& Podolak}{Vazan
  et~al.}{2014}]{Vazan14}
Vazan A.,  Helled R.,  Kovetz A.,    Podolak M.,  2014, submitted, 0, 0

\bibitem[\protect\citeauthoryear{von Zahn, Hunten \& Lehmacher}{von Zahn
  et~al.}{1998}]{Zahn98}
von Zahn U.,  Hunten D.~M.,    Lehmacher G.,  1998, J. Geophys. Res., 103,
  22815

\bibitem[\protect\citeauthoryear{Walin}{Walin}{1964}]{Walin64}
Walin G.,  1964, Tellus XVI, 3, 389

\bibitem[\protect\citeauthoryear{Wang, Ayala, Kasprzak \& Grabowski}{Wang
  et~al.}{2005}]{Wang05}
Wang L.,  Ayala O.,  Kasprzak S.,    Grabowski W.,  2005, J. Atmosph. Sci., 62,
  2433

\bibitem[\protect\citeauthoryear{Weir, Mitchell \& Nellies}{Weir
  et~al.}{1996}]{Weir96}
Weir S.~T.,  Mitchell A.~C.,    Nellies W.~J.,  1996, Phys. Rev. Lett., 76,
  1860

\bibitem[\protect\citeauthoryear{Wilson \& Militzer}{Wilson \&
  Militzer}{2010}]{WilMil10}
Wilson H.~F.,  Militzer B.,  2010, PRL, 104, 121101

\bibitem[\protect\citeauthoryear{Wood, Garaud \& Stellmach}{Wood
  et~al.}{2013}]{Wood13}
Wood T.~S.,  Garaud P.,    Stellmach S.,  2013, ApJ, 768, 157

\end{thebibliography}


\bsp

\label{lastpage}
\end{document}